\documentclass[fleqn,usenatbib,onecolumn]{mnras}

\usepackage[T1]{fontenc}

\DeclareRobustCommand{\VAN}[3]{#2}
\let\VANthebibliography\thebibliography
\def\thebibliography{\DeclareRobustCommand{\VAN}[3]{##3}\VANthebibliography}

\usepackage{graphicx}  

\usepackage{adjustbox}

\usepackage{amsmath}   
\usepackage[left]{lineno}
\usepackage{epstopdf}
\usepackage{ulem}
\usepackage{amssymb}
\usepackage{times}
\usepackage[usenames,dvipsnames]{pstricks}
\usepackage{psfrag}
\usepackage{lipsum}
\usepackage{natbib}
\usepackage{rotating}
\usepackage{hyperref}

\usepackage{newtxtext,newtxmath}

\def\araa{ARA\&A}
\def\apj{ApJ}
\def\apjl{ApJ}
\def\apjs{ApJS}
\def\ao{Appl.~Opt.}
\def\apss{Ap\&SS}
\def\aap{A\&A}
\def\aaps{A\&AS}
\def\mnras{MNRAS}
\def\prl{Phys.~Rev.~Lett.}
\def\nat{Nature}

\newcommand{\be}{\begin{equation}}
\newcommand{\ee}{\end{equation}}
\newcommand{\bary}{\begin{eqnarray}}
\newcommand{\eary}{\end{eqnarray}}

\title[Quasi-spherical Outflow in a stratified medium]{Late-afterglow Emission from a Quasi-spherical Outflow in a stratified environment}

\author[Fraija et al.]{
N.~ Fraija,$^{1}$\thanks{E-mail: nifraija@astro.unam.mx}
B.~ Betancourt Kamenetskaia,$^{2}$
A.~ Galv\'{a}n,$^{3}$
A.~ Montalvo,$^{1}$
A.~ C.~ Caligula Do E.~ S.~ Pedreira,$^{1}$
\newauthor
P.~ Veres,$^{4}$
R.~L.~ Becerra,$^{5}$
M.G. Dainotti,$^{6,7}$
S.~ Dichiara,$^{8}$
H.~ Le\'{o}n Vargas $^{3}$
\\
$^{1}$Instituto de Astronom\'ia, Universidad Nacional Aut\'{o}noma de M\'{e}xico, A. P. 70-264, Cd. Universitaria, Ciudad de M\'{e}xico, M\'{e}xico\\
$^{2}$Cosmology, Gravity, and Astroparticle Physics Group, Center for Theoretical Physics of the Universe,
Institute for Basic Science (IBS), Daejeon, 34126, Korea\\
$^{3}$Instituto de F\'{i}sica, Universidad Nacional Aut\'{o}noma de M\'{e}xico, Apdo. Postal 70-264, Cd. Universitaria, Ciudad de M\'{e}xico 04510\\
$^{4}$Center for Space Plasma and Aeronomic Research (CSPAR), University of Alabama in Huntsville, Huntsville, AL 35899, USA\\
$^{5}$Department of Physics, University of Rome - Tor Vergata, via della Ricerca Scientifica 1, 00100 Rome, IT\\
$^{6}$Division of Science, National Astronomical Observatory of Japan, 2-21-1 Osawa, Mitaka, Tokyo 181-8588, Japan\\
$^{7}$The Graduate University for Advanced Studies (SOKENDAI), 2-21-1 Osawa, Mitaka, Tokyo 181-8588, Japan\\
$^{8}$Department of Astronomy and Astrophysics, The Pennsylvania State University, 525 Davey Lab, University Park, PA 16802, USA\\
}

\date{Accepted XXX. Received YYY; in original form ZZZ}

\pubyear{\the\year{}}

\begin{document}
\label{firstpage}
\pagerange{\pageref{firstpage}--\pageref{lastpage}}
\maketitle

\begin{abstract}
Gamma-ray bursts (GRBs) are cosmic events occurring at large distances beyond our galaxy. They provide a unique opportunity to study electromagnetic patterns not seen elsewhere. When the collimated GRB outflow interacts with the outer layers of a star or the wind generated by a binary neutron star merger, it releases energy, forming a quasi-spherical outflow around it. This broad outflow begins to radiate once it has transferred enough energy to the surrounding medium. We have developed a new analytical model that describes the synchrotron afterglow scenario of the quasi-spherical outflow, including factors such as stratified density, self-absorption regime, and the fraction of electrons accelerated by the shock front. We also successfully describe the multiwavelength observations of a sample of llGRB afterglows (GRB 980425, 031203, 060218, 100316D, 130603B, 150101B and 171205A) that exhibited a late component, analyzed in both stellar wind and constant-density environments. Our analysis shows that a constant-density environment is favored. Additionally,  we consider the multiwavelength upper limits of the short bursts reported in the Swift-BAT database.

\end{abstract}



\section{Introduction}
The most luminous events in the Universe are gamma-ray bursts (GRBs), which are created when massive stars collapse \citep{1993ApJ...405..273W,1998ApJ...494L..45P, Woosley2006ARA&A,Cano2017} or when two compact objects like a neutron star (NS) and a black hole \citep[BHs,][]{1992ApJ...395L..83N} or two NSs \citep{1992ApJ...392L...9D, 1992Natur.357..472U, 1994MNRAS.270..480T, 2011MNRAS.413.2031M} merge, as exhibited by the quasi-simultaneous  gravitational-wave (GW) and electromagnetic  detections in GW170817/GRB170817A \citep{PhysRevLett.119.161101}.   According to the duration of the main episode ($T_{90}$),\footnote{From 5\% to 95\% of the total counts observed from a burst's prompt emission occur within a time interval denoted by $T_{90}$.} usually described by an empirical ``Band" function \citep{1993ApJ...413..281B}, GRBs are further divided into two categories: short GRBs (sGRBs) with $T_{90}\leq 2\mathrm{\,s}$ and long GRBs (lGRBs) with  $T_{90} \ge 2\mathrm{\,s}$ \citep{mazets1981catalog, kouveliotou1993identification}.  SGRBs are the outcome of the merger of compact objects, whereas lGRBs are the product of the collapse of massive stars. Irrespective of the progenitor, a relativistic collimated jet is launched producing the prompt GRB emission via internal shocks or magnetic reconnections, and subsequently afterglow radiation when it interacts with the external environment. However, during the core-collapses of dying massive stars or the coalescence of compact objects, a cocoon / shock breakout emission could also be expected  \cite[e.g., see][]{2014ApJ...784L..28N, 2014ApJ...788L...8M,2017ApJ...848L...6L, 2018PhRvL.120x1103L, 2017ApJ...834...28N,2018MNRAS.473..576G}.

As the GRB jet makes its way through the stellar envelope of an enveloped star or as it traverses inside the neutrino/magnetically-driven wind previously ejected during the binary NS merger, the energy deposited laterally is significant and, therefore, will form a cocoon that engulfs the jet  \citep{2002MNRAS.337.1349R, 2006ApJ...652..482P,2014ApJ...788L...8M, 2014ApJ...784L..28N}. The luminosity of the jet during the propagation phase should be similar to that seen during the prompt GRB phase. This indicates that the cocoon may be provided with an amount of energy that is similar to that of a prompt emission.  \cite{2014ApJ...784L..28N} numerically showed that when a low-luminosity jet is considered, a hot cocoon confining the jet is formed.  \cite{2014ApJ...788L...8M} found that when the jet has low or high luminosity a weak cocoon emission is expected from a typical sGRB. As soon as the relativistic jet reaches the outermost material, the cocoon emerges with a wider angle than the GRB jet and expands along the jet's axis. Beyond the outermost material, the external pressure decreases abruptly, so the cocoon can accelerate and expand until it becomes transparent.  Accelerated material from the cocoon fireball will continue moving in the jet's axis.  The cocoon fireball with a typical ejected mass in the range of $\sim10^{-6} - 10^{-4}\,{\rm M}_\odot$ can be modelled as a quasi-spherical outflow with Lorentz factor $\Gamma\sim2 - 3 $  \citep{2014ApJ...788L...8M, 2014ApJ...784L..28N, 2015ApJ...802...95R} once it emerges from the breakout, although in some scenarios the Lorentz factor could achieve values of $\sim 10$ \citep{2017ApJ...834...28N}. The quasi-spherical material interacting with the external medium could contribute to the electromagnetic emission in several frequencies and temporal scales, so afterglow measurements may help us learn about the progenitor and constrain the density of the external environment \citep{2016ApJ...818..190F, 2016ApJ...831...22F}.


 The first detectable electromagnetic signature from a star explosion might occur when a shock breaks out and travels outward through the stellar envelope. Before it breaks forth, radiation dominates the shock, and after, when the optical depth becomes unity, the photons in the trough layer all leave at once, producing an intense and brief electromagnetic signature that goes from UV to gamma rays. Photons in the UV band are expected in the Newtonian regime \citep{1974ApJ...187..333C, 1978ApJ...225L.133F, 1978ApJ...223L.109K, 1981Ap&SS..78..105I,  10.1046/j.1365-8711.2000.03922.x, 1992ApJ...393..742E, 2012ApJ...747..147K}, and  up to gamma-rays are expected in the relativistic regime \citep{2010ApJ...716..781K, 2010ApJ...725..904N, 2012ApJ...747...88N}.  Once this shock-breakout material makes contact with the circumstellar medium, it will be powerless to affect the launching phase of the expansion until the quantity of material that is swept up is equal to the mass that was ejected. At this point, the ejecta masses had already started to gradually slow down and take on a velocity structure, with the material moving more quickly at the front of the ejected mass than at the back.

In several circumstances, the density profile of the circumburst medium around GRBs in the form $\propto r^{-k}$ with $k=0$ (ISM) and $k=2$ (stellar wind) has been used to describe the multiwavelength afterglow observations \cite[e.g., see][]{2013ApJ...776..120Y, 2013ApJ...774...13L, 2008Sci...321..376K, 2013ApJ...779L...1K, 2020ApJ...900..176L, 2017ApJ...848...15F, 2019ApJ...879L..26F}.  For instance, \cite{2013ApJ...776..120Y} looked at over a dozen GRBs and came to the conclusion that the circumburst environment was not homogenous nor a stellar-wind medium, but rather a mixture of the two, whose density-profile index falls between $0.4$ and $1.4$.  After analyzing a bigger sample of 146 GRBs, \cite{2013ApJ...774...13L} reported a  preferred index around ${ k\sim1}$. Furthermore, \cite{2008Sci...321..376K} found a density-profile index of $k=2.5$  after investigating the possibility that accretion of a stellar envelope in a BH could explain the plateau phase found in some X-ray light curves.   On the other hand, \cite{2012ApJ...747..118M} demonstrated that the spectrum variety of Type II bright supernovae (SN IIa) may be partially explained by variations in the density slope of the dense wind caused by non-steady mass loss.  They advocated for a $\propto r^{-k}$ wind density structure and found that variations in spectral SN development are substantially dependent on the ${ k}$ index.   \cite{2013ApJ...779L...1K} performed a multi-wavelength analysis and modeled the GRB 130427A observations from radio to GeV energy bands with the standard synchrotron model. They reported that the multiwavelength afterglow observations were consistent with a density profile index  between ${\rm k}=0$ and ${\rm k}=2$.  SN 2020bvc is a recent case study into the possible relationship between SN and GRB emission and a stratified environment \citep{2020A&A...639L..11I}.  The X-ray light curve is in good agreement with expectations for a decelerated material in a stratified medium with ${\rm k=1.5}$, which is consistent with the GRB-associated, broad-lined Ic SN 1998bw. Several bursts of multiwavelength data have shown an afterglow transition from stellar wind environment to ISM \citep[e.g., see][]{2003ApJ...591L..21D, 2007ApJ...664L...5K, 2009MNRAS.400.1829J, 2020ApJ...900..176L, 2017ApJ...848...15F, 2019ApJ...879L..26F}.\\ 

A large half-opening angle, the essential parameter for characterizing the cone-shaped area from which the relativistic outflow of a GRB originates, and the low luminosity in the gamma-ray band are two factors that have led some authors to classify GRBs as low-luminosity (ll)GRBs \citep{2011ApJ...739L..55B, 2013RSPTA.37120275H, 2014A&A...566A.102S, 2017A&A...605A.107C}.
Similarly, llGRBs are associated with the shock breakout and cocoon materials \citep{2011ApJ...739L..55B, 2014ApJ...788L...8M, 2015MNRAS.448..417B}. Because we consider the synchrotron emission of an outflow with a large opening angle (quasi-spherical), we apply this scenario to bursts with low luminosity. So far, there are five confirmed llGRB:  GRB 980425 \citep{1998Natur.395..670G}, GRB 031203 \citep{2004A&A...419L..21T}, GRB 060218 \citep{2006Natur.442.1008C}, GRB 100316D \citep{2011ApJ...740...41C}
and GRB 171205A \citep{2019Natur.565..324I}.  Synchrotron afterglow scenarios from a quasi-spherical material used to describe some llGRBs assumed either a constant density medium ($k=0$) or a stellar wind ($k=2$) as the circumburst medium, without considering the self-absorption regime, or that the fraction of electrons accelerated in the shock front could be less than unity.  For instance,~\cite{2015MNRAS.448..417B} considered a wind shock-breakout model to fit the radio and X-ray data of GRB 980425 introducing two power-law electron distributions; one distribution for the radio wavelengths and another for X-ray data. In addition, they explained the prompt emission of GRB 060218 as arising from a mildly relativistic shock breakout (with a low value of Lorentz factor $\sim 1.3$), and its radio afterglow from the interaction between ejecta and the surrounding wind medium. The authors suggested that the X-ray and radio emission might originate from different electron populations or emission processes. To overcome this limitation, we have systematically generalized the afterglow model to a stratified medium with a general index $0<k<3$, allowing for the analysis of environments beyond ISM or wind, such as those predicted for merger ejecta or progenitors with non-steady mass loss \citep{2005ApJ...631..435R, 2012ApJ...747..118M}.  Furthermore, we extend the synchrotron spectra to include both the coasting phase as well as the fraction of electrons accelerated by the shock front and self-absorption regime, crucial for modeling early radio emission \citep{2019ApJ...884L..58U} and to unify the relativistic and sub-relativistic regimes, enabling continuous modeling from prompt cocoon or breakout material to late-time emission.




\vspace{3cm}





In this paper, we extend the analytical synchrotron afterglow scenario of the quasi-spherical outflow model that was previously utilized to characterize the earlier X-ray and radio afterglow data in GRB 170817A \cite[see][]{2019ApJ...884...71F}.  Firstly, we assume that the external medium surrounding the burst is stratified, with a profile density of $\propto r^{-k}$ with a corresponding value of ${\rm k}$ between $0\leq k < 3$. Secondly, we extend the synchrotron scenario to the self-absorption regime. Thirdly, we derive the coasting phase when the quasi-spherical outflow is not yet decelerated, and finally, we consider that only a fraction of electrons are accelerated by the shock front. We use this model to interpret a sample of llGRBs, bursts with late observations. We consider the database of the Burst Area Telescope (BAT) instrument on board the Neil Gehrels \textit{Swift} with sGRBs located between 100 and 200 Mpc.   This paper is arranged as follows: In Section \S\ref{sec2} we derive the synchrotron scenario of the quasi-spherical outflow decelerated in a stratified circumburst medium. In Section \S\ref{sec3}, we apply the proposed analytical model to describe the multi-wavelength observations of a sample of bursts and provide constraints to other ones. In Section \S\ref{sec5}, we synthesize prior research and clarify distinctions with the current scenario. Finally, in Section \S\ref{sec4}, we provide a summary of our work and some concluding remarks. 

\section{Quasi-spherical outflow in a stratified medium}
\label{sec2}
During the initial coasting phase, the relativistic outflow is not impacted by the circumburst medium. Hence, in the coasting phase, the bulk Lorentz factor remains constant $\Gamma = \Gamma_0$ and the radius develops as $r=\frac{c\beta_0 t}{(1+z)}$ with $\beta_0=\sqrt{\Gamma^2_0-1}/\Gamma_0$. After the coasting phase, the deceleration phase begins. Here the ejected mass gains a velocity structure, with the matter in the front of the expelled mass moving faster than the one in the rear \citep{2000ApJ...535L..33S}. The quasi-spherical material then transmits a considerable portion of its energy to its surrounding environment, leading to a forward shock.
\cite{2001ApJ...551..946T} investigated the ejected mass's acceleration at relativistic and sub-relativistic velocities.  They discovered that at the sub- and ultra-relativistic limit, the isotropic-equivalent kinetic energy may be represented as a power-law velocity distribution provided by 

{\small
\be\label{E_beta}
E (\beta \Gamma)= \tilde{E}\,\left[ (\Gamma\beta)^{-(5.35-\frac{2}{\gamma_p})\frac{1}{q} } + (\Gamma\beta)^{-(1.58-\frac{1}{\gamma_p})\frac{1}{q}}\right]^{q\gamma_p}\,,
\ee
}
where $\gamma_p=1+1/n_p$ with $n_p=3$ the polytropic index,  $\tilde{E}$ is the fiducial energy, $\beta=\sqrt{\Gamma^2-1}/\Gamma$ and $q=4.1$ is a parameter \citep[e.g., see][]{2001ApJ...551..946T}.  In the limiting cases, Eq. \ref{E_beta} becomes $E \simeq \tilde{E}\, (\Gamma\beta)^{-(1.58\gamma_p-1)}$ at the relativistic velocities ($\beta\Gamma\gg 1$), and $E\simeq \tilde{E}\, (\Gamma\beta)^{-(5.35\gamma_p-2)}$ at the sub-relativistic velocities ($\Gamma \beta\ll 1$). Additionally,  we consider that a quasi-spherical material is decelerated in a stratified density profile given by $n(r)=n_{\rm k}\left(\frac{r}{r_0}\right)^{-k}$ with $0\leq k <3$.  For instance, with a density profile index of ${ k=2}$,  $n_{\rm k}r^{k}_0=A_{st\rm }\,3\times 10^{35}\,{\rm cm^{-1}}$, with $A_{\rm st}$ the parameter of density profile.  Hereafter, we refer to quantities in the observer and  comoving frames with unprimed and primed terms, respectively.  We also adopt the convention $Q_{\rm x}=Q/10^{\rm x}$ in c.g.s. units.

\subsection{Mildly- relativistic regime}
At the relativistic velocities $E \simeq \tilde{E}\, (\Gamma\beta)^{-\alpha_s}$ with $1.1\leq\alpha_s\leq2.1$,  the adiabatic evolution of the forward shock with an isotropic equivalent-kinetic energy  becomes $\tilde{E}\,\left(\beta\Gamma\right)^{-\alpha_s}=\frac{4\pi}{3} m_pc^2 n(r) r^{3} \beta^2 \Gamma^2$ \citep[Blandford-McKee solution;][]{1976PhFl...19.1130B}.  Given a radial distance $r=\frac{2c\beta \Gamma^2 t}{(1+z)}$, the bulk Lorentz factor evolves as 
\be\label{Gamma}
\Gamma\propto  \left(1+z\right)^{-\frac{k-3}{\alpha_s+8-2k}}\, A^{-\frac{1}{\alpha_s+8-2k}}_{\rm st} \,  \beta^{-\frac{\alpha_s+5-k}{\alpha_s+8-2k}}\, \tilde{E}^{\frac{1}{\alpha_s+8-2k}}  \,t^{\frac{k-3}{\alpha_s+8-2k}} \,,
\ee
where the deceleration time evolves as $t_{\rm dec}\propto  \left(1+z\right)\, A^{\frac{1}{k-3}}_{\rm st} \, \beta^{\frac{\alpha_s+5-k}{k-3}} \tilde{E}^{-\frac{1}{k-3}}  \,\Gamma^{\frac{\alpha_s+8-2k}{k-3}}$.  When the quasi-spherical material enters the post-jet-break decay phase ($\Gamma\simeq 1/\theta_c$, with $\theta_c$ the opening angle), the bulk Lorentz factor evolves as 
\be\label{Gamma_l_ism}
\Gamma \propto (1+z)^{-\frac{k-3}{\alpha_s+6-2k}} A^{-\frac{1}{\alpha_s+6-2k}}_{\rm st}\beta^{-\frac{\alpha_s+5-k}{\alpha_s+6-2k}}\tilde{E}^{\frac{1}{\alpha_s+6-2k}}t^{\frac{k-3}{\alpha_s+6-2k}} \,,
\ee
where the break time becomes $t_{\rm br} \propto  \left(1+z\right)\, A^{\frac{1}{k-3}}_{\rm st} \,\beta^{\frac{\alpha_s+5-k}{k-3}}  \tilde{E}^{-\frac{1}{k-3}}  \,\theta_c^{-\frac{\alpha_s+8-2k}{k-3}}$.  Given the non evolution of bulk Lorentz factor ($\Gamma \propto t^0$) during the coasting phase, the evolution of the bulk Lorentz factor can be summarized as 
{\small
\begin{eqnarray}
\label{gammas}
\Gamma \propto 
\begin{cases} 
t^0,\hspace{2.1cm}  t < t_{\rm dec} , \cr
t^{\frac{k-3}{\alpha_s+8-2k}},\hspace{1.1cm}  t_{\rm dec} \leq t\leq t_{\rm br}, \cr
t^{\frac{k-3}{\alpha_s+6-2k}}\, \hspace{1.2cm} t_{\rm br} \leq t.\hspace{.2cm}\cr
\end{cases}
\end{eqnarray}
}
For the case of $\alpha_s=0$ and $k=0$, the isotropic equivalent-kinetic energy is constant, and the evolution of the bulk Lorentz factor $\Gamma\propto t^0$, $\propto t^{-\frac{3}{8}}$, and $\propto t^{-\frac{1}{2}}$ are obtained \citep{1999A&AS..138..537S,1998ApJ...497L..17S,1999ApJ...519L..17S}.\\

In forward-shock models, the arrangement of the accelerated electrons depends on their Lorentz factors ($\gamma_e$), and the electron power index $p$. They follow the distribution $N(\gamma_e)\,d\gamma_e \propto \gamma_e^{-p}\,d\gamma_e$ for $\gamma_m\leq \gamma_{\rm e}$, where $\gamma_m= \frac{m_{\rm p}}{m_{\rm e}}g(p)\varepsilon_{\rm e}(\Gamma-1)\zeta^{-1}_e$ is the minimum electron Lorentz factor, $m_{\rm p}$ and $m_{\rm e}$ the proton and electron mass, respectively, $\varepsilon_{\rm e}$ the fraction of energy given to accelerate electrons, $\zeta_{e}$ the fraction of electrons that were accelerated by the shock front \citep{2006MNRAS.369..197F} and $g(p)=\frac{p-2}{p-1}$. The energy density $U=[(\hat\gamma\Gamma +1)/(\hat\gamma - 1)](\Gamma -1)n(r) m_pc^2$ with $\hat\gamma$ the adiabatic index \citep{1999MNRAS.309..513H} and its respective fraction given to magnetic field ($\varepsilon_B$) are used to calculate the intensity of the blastwave's comoving magnetic field $B'^2/(8\pi)=\varepsilon_B U$. The cooling electron Lorentz factor is {\small $\gamma_{\rm c} = \frac{6\pi m_e c}{\sigma_T}(1+Y)^{-1}\Gamma^{-1}B'^{-2}t^{-1}$}, where $\sigma_T$ is the cross-section in the Thomson regime, $c$ is the speed of light  and $Y$ is the Compton parameter \citep{2001ApJ...548..787S, 2010ApJ...712.1232W}.  The synchrotron spectral breaks and the synchrotron radiation power per electron in the comoving frame are given by $\nu_{\rm i}=\frac{q_e}{2\pi m_ec}(1+z)^{-1}\Gamma\gamma^{2}_{\rm i}B'$ (with ${\rm i=m}$ and ${\rm c}$) and $P_{\nu_m}\simeq \frac{\sqrt{3}q_e^3}{m_ec^2} (1+z)^{-1} \Gamma\, B'$, respectively, with $q_e$ the elementary charge \citep[e.g., see][]{1998ApJ...497L..17S, 2015ApJ...804..105F}. The synchrotron spectral breaks in the  self-absorption regime are derived from  $\nu_{\rm a,1}=\nu_{\rm c}\tau^{\frac35}_{0,m}$,  $\nu_{\rm a,2}=\nu_{\rm m}\tau^{\frac{2}{p+4}}_{0,m}$ and $\nu_{\rm a,3}=\nu_{\rm m}\tau^{\frac35}_{0,c}$ with the optical depth given by $\tau_{0,i}\simeq\frac{5}{3-k}\frac{q_en(r)r}{B'\gamma^5_{\rm i}}$ \citep{1998ApJ...501..772P}.   Considering  the total number of emitting electrons $N_e= n(r) \frac{4\pi}{3-k} r^3$, the radiation power and the synchrotron spectral breaks, the maximum flux becomes {\small $F_{\rm \nu, max}=\frac{(1+z)^2}{4\pi d_z^2}N_eP_{\nu_m}$},  where {\small $d_{\rm z}=(1+z)\frac{c}{H_0}\int^z_0\,\frac{d\tilde{z}}{\sqrt{\Omega_{\rm M}(1+\tilde{z})^3+\Omega_\Lambda}}$}  \citep{1972gcpa.book.....W}  is the luminosity distance.   We assume for the cosmological constants a spatially flat universe $\Lambda$CDM model with  $H_0=69.6\,{\rm km\,s^{-1}\,Mpc^{-1}}$, $\Omega_{\rm M}=0.286$ and  $\Omega_\Lambda=0.714$ \citep{2016A&A...594A..13P}.

\subsection{Sub-relativistic regime}

During this regime, the isotropic equivalent-kinetic energy becomes $E \simeq \tilde{E}\, \beta^{-\alpha_s}$ with $3\leq\alpha_s\leq5.2$, and the dynamics of the non-relativistic ejecta mass is described by the Sedov-Taylor solution.   Then, given the evolution of the radius $r\propto\,\left({1+z}\right)^{-\frac{\alpha_s+2}{\alpha_s+5-k}}\,A^{-\frac{1}{\alpha_s+5-k}}_{\rm st}\, \tilde{E}^{\frac{1}{\alpha_s+5-k}}\,t^{\frac{\alpha_s+2}{\alpha_s+5-k}}$,  the velocity can be written as 
{\small
\be\label{beta_dec}
\beta \propto \,\left(1+z\right)^{\frac{3-k}{\alpha_s+5-k}}\,A^{-\frac{1}{\alpha_s+5-k}}_{\rm st}\,\tilde{E}^{\frac{1}{\alpha_s+5-k}}\, t^{\frac{k-3}{\alpha_s+5-k}}\,,
\ee
}
where the deceleration time evolves as $t_{\rm dec}\propto \beta^{\frac{\alpha_s+5-k}{k-3}}$.  During the coasting phase, the velocity of the quasi-spherical material is not altered and therefore becomes constant ($\beta\propto t^0$). Taking into account the coasting and deceleration phase, the evolution of the velocity can be summarized as

{\small
\begin{eqnarray}
\label{gammas1}
\beta \propto 
\begin{cases}
t^0,\hspace{2.1cm}  t < t_{\rm dec} , \cr
t^{\frac{k-3}{\alpha_s+5-k}},\hspace{1.3cm}  t_{\rm dec} \leq t\,. \cr
\end{cases}
\end{eqnarray}
}
For the case of $\alpha_s=0$ and $k=0$, the evolution of the velocity in \cite{2013ApJ...771...54S} is recovered.\\ 


For the sub-relativistic regime, we assume that the shocked-accelerated electrons have the same distribution as the previous relativistic regime $\frac{dn}{d\gamma_e}\propto \gamma_e^{-p}$ for $\gamma_{\rm e}\geq \gamma_{\rm m}$,  with {\small $\gamma_{\rm m}\propto  \,\beta^2$} the Lorentz factor of the lowest-energy electrons. In this case, the magnetic field evolves as {\small $B'\propto A^\frac12_{\rm st}$ $\beta^{\frac{2-k}{2}}\,t^{-\frac{k}{2}}$} and the Lorentz factor of the highest energy electrons, which are efficiently cooled by this process as {\small $\gamma_{\rm c}\propto\,A^{-1}_{\rm st} \beta^{k-2}\,t^{k-1}$}.  The synchrotron spectral (characteristic and cooling) breaks and the synchrotron radiation power per electron evolve as {\small $\nu_{\rm m}\propto A^\frac{1}{2}_{\rm st} \beta^\frac{10-k}{2}\,t^{-\frac{k}{2}}$},  {\small $\nu_{\rm c}\propto A^{-\frac32}_{\rm st} \beta^{\frac{3k-6}{2}}\,t^{\frac{3k-4}{2}}$} and {\small $P_{\rm \nu, max} \propto \,A^\frac12_{\rm st} \beta^{\frac{2-k}{2}}\,t^{-\frac{k}{2}}$}, respectively. Considering the total number of emitting electrons {\small $N_{\rm e}\propto A_{\rm st}\beta^{3-k}\, t^{3-k}$}, the radiation power, and the  synchrotron spectral breaks, the maximum flux varies as {\small $F_{\rm \nu,max}\propto  A^{\frac32}_{\rm st}\, \beta^{\frac{8-3k}{2}}\,t^{\frac{3(2-k)}{2}}$}.  In the self-absorption regime, the synchrotron spectral breaks evolve as  {\small $\nu_{\rm a,1}\propto A^{\frac45}_{\rm st} \beta^{-\frac{4k+5}{5}}\,t^{\frac{3-4k}{5}}$},  {\small $\nu_{\rm a,2}\propto  A^{\frac{p+6}{2(p+4)}}_{\rm st}\beta^{\frac{10p-kp-6k}{2(p+4)} }\,t^{\frac{4-kp-6k}{2(p+4)}}$}, and {\small $\nu_{\rm a,3}\propto  A^{\frac95}_{\rm st} \beta^{\frac{15-9k}{5}}\,t^{\frac{8-9k}{5}}$} for {\small $\nu_{\rm a,1}\leq \nu_{\rm m} \leq \nu_{\rm c}$},
{\small $\nu_{\rm m} \leq\nu_{\rm a,2}\leq \nu_{\rm c}$ and {\small $\nu_{\rm a,3}\leq \nu_{\rm c} \leq  \nu_{\rm m}$}, respectively \citep[details of the derivation are explicitly written in][]{2021ApJ...907...78F}.\\

\subsection{Analysis and Description}

Appendix~\ref{Appendix} summarizes our analytical synchrotron forward-shock model from the deceleration of a quasi-spherical material in a stratified environment during the relativistic and sub-relativistic regimes. In addition, we include the evolution of the quasi-spherical material in the coasting and post-jet break decay stages.  Table~\ref{Table2} and Table~\ref{Table3} summarize, for each cooling condition, the evolution of the synchrotron fluxes and the closure relations  respectively.

Table \ref{TableDensityParameter} shows the evolution of the density parameter in each cooling condition of the synchrotron afterglow model from a quasi-spherical material in both the relativistic and sub-relativistic regimes. For each regime, the coasting, deceleration and post-jet break decay phases are exhibited. For instance, 
in the sub-relativistic regime the synchrotron flux as a function of the density parameter is given by $F_{\nu}:~\propto A_{st}^{\frac{4\alpha_s+13)}{3(\alpha_s+5-k)}}$ for $\nu_{\rm a,1}< \nu <\nu_{\rm m}$, $\propto A_{st}^{\frac{19+p(\alpha_s-5\alpha_s)+5}{4(\alpha_s+5-k)}}$ for $\nu_{\rm m}< \nu < \nu_{\rm c}$ and $A_{st}^{\frac{p(\alpha_s-5) + 2(\alpha_s+5))}{4(\alpha_s+5-k)}}$ for $\nu_{\rm c}< \nu$, and in the relativistic regime as $F_{\nu}:~\propto A_{st}^{\frac{4(\alpha_s+3)}{3(\alpha_s+8-2k)}}$ for $\nu_{\rm a,1}< \nu <\nu_{\rm m}$, $\propto A_{st}^{\frac{16+\alpha_s(p+5)}{4(\alpha_s+8-2k)}}$ for $\nu_{\rm m}< \nu < \nu_{\rm c}$ and $A_{st}^{\frac{\alpha_s(p+2)}{4(\alpha_s+8-2k)}}$ for $\nu_{\rm c}< \nu$.  On the other hand, the synchrotron light curve after the jet break  as a function of the density parameter is given by $F_{\nu}:~\propto A_{st}^{\frac{4(\alpha_s+1)}{3(\alpha_s+6-2k)}}$ for $\nu <\nu_{\rm m}$, $\propto A_{st}^{\frac{6+5\alpha_s+p(\alpha_s-2)}{4(\alpha_s+6-2k)}}$ for $\nu_{\rm m}< \nu < \nu_{\rm c}$ and $A_{st}^{\frac{(p+2)(\alpha_s-2)}{4(\alpha_s+6-2k)}}$ for $\nu_{\rm c}< \nu$. These relations show that any variation of the density parameter, in both the relativistic and lateral expansion phases, will be more pronounced in the first two cases, that is in $\nu <\nu_{\rm m}$ and $\nu_{\rm m}< \nu < \nu_{\rm c}$. This points out that this variation will be enhanced in low-energy bands, namely radio and optical. It is also important to note that this feature will be intensified as the value of the velocity distribution parameter $\alpha_{\rm s}$ is increased, as well as that of the stratification parameter $k$. Given this discussion, a transition between different circumburst density parameters will be more efficiently observed in the radio and optical bands with high values of stratification and of the velocity distribution parameter. On the other hand, this feature will also vary between both phases. In the relativistic phase, the second regime ($\nu_{\rm m}< \nu < \nu_{\rm c}$) will enhance the variation of the density, while in the lateral expansion phase, it is the first regime ($\nu <\nu_{\rm m}$) that amplifies this difference. Therefore, it is expected that the contrast will be more easily observed in the optical band during the relativistic phase, while in the lateral expansion phase, this will be simpler in lower energy bands, such as radio.\\

The synchrotron spectral breaks during the deceleration phase evolve as $\nu_{\rm m}\propto t^{-\frac{30 + k(\alpha_s-8)}{2(\alpha_s+5-k)}}$  and  $\nu_{\rm c}\propto t^{\frac{k(4+3\alpha_s) - 4\alpha-2}{2(\alpha_s+5-k)}}$ for the non-relativistic regime, and as  $\nu_{\rm m}\propto t^{-\frac{24+k(\alpha_s-6)}{2(\alpha_s+8-2k)}}$ and  $\nu_{\rm c}\propto t^{-\frac{(\alpha_s+2)(4-3k)}{2(\alpha_s+8-2k)}}$ for the relativistic regime. In a similar way to the work by \cite{1999ApJ...524L..47G} for the relativistic regime using the standard synchrotron afterglow model, we propose the evolution of the estimated flux a useful tool for locating the synchtrotron emission from the non-relativistic and relativistic quasi-spherical material.  \cite{1999ApJ...524L..47G} studied the evolution of spectral indexes and break energy of the prompt episode ($T_{90}\simeq 40\,{\rm s}$) followed by a smooth emission tail that lasted $\sim 400\,{\rm s}$ observed in GRB 980923. They found that the low-energy spectral index in the smooth tail evolved as the synchrotron cooling break  $\propto t^{-0.52\pm0.12}$, thus concluding that the forward-shock had started during the prompt episode. Afterwards, spectral analyses of burst tails were performed  to identify early afterglow emissions \citep[e.g., see][]{2005ApJ...635L.133B, 2006MNRAS.369..311Y}.

\section{Application}\label{sec3}

\subsection{A sample of LLGRBs: GRB 980425, 031203, 060218, 100316D, 130603B, 150101B and 171205A}

\subsubsection{GRB 980425}
GRB 980425 was detected on April 25, 1998 by the BATSE instrument on board the \textit{Compton Gamma Ray Observatory} (CGRO) and both by the Gamma-ray Burst Monitor (GRBM) and the second Wide Field Cameras (WFC) on board the \textit{BeppoSAX}, respectively. \citep{1998GCN67,IAU1998}. The duration of the burst is about 30~s. Three days after the trigger, the \textit{Australian Telescope Compact Array} (ATCA) found three radio sources within the 8 arcmin error box. The brightest source was located at coordinates RA(J2000)=19h 35m 03.31s and Dec(J2000)$=-52^{\circ} 50^{\prime} 44.7^{\prime \prime}$, which coincided with the position of SN 1998bw \citep{1998GCN63}.   This association with the SN host galaxy ESO 184-G82 and spectroscopy led to the determination of the burst redshift $z=0.0085\pm0.0002$ \citep{IAU19986896}.  The isotropic-equivalent energy in the $1 - 10^4\,{\rm keV}$ energy range was $E_{\rm \gamma, iso}=(9.29\pm0.35)\times 10^{47}\,{\rm erg}$.  The corresponding total fluence in the gamma-ray (30 - 400 keV) and X-ray (2 - 30 keV) bands were $3.40\times10^{-6}\,{\rm erg\,cm^{-2}}$ and  $1.99\times10^{-6}\,{\rm erg\,cm^{-2}}$, respectively. The radio observations at 20, 13, 6 and 3 cm began to be collected with ATCA 3 days after the trigger time. These radio observations were performed during almost 81 days.

\subsubsection{GRB 031203}

On December 03 2003 at 22:01:28 UTC, GRB 031203 was detected by the {\itshape INTEGRAL}/IBIS instrument at coordinates RA(J2000)$=08^{\rm h} 02^{\rm m} 30^{\rm s}$ and Dec(J2000)$=-39^{\circ} 50^{\prime} 49^{\prime \prime}$ \citep{2003GCN..2459....1G} with a duration T90 in the $15-200$ keV band of $\sim 20$~ s. Subsequent follow-up observations were carried out by the \textit{SMARTS} telescope \citep{2003GCN..2463....1B, 2003GCN..2464....1S}, the Newton X-ray Multiple Mirror \textit{XMM-Newton}  \citep{2003GCN..2473....1F}, the Advanced CCD Imaging Spectrometer (ACIS) instrument onboard the Chandra X-ray Observatory (CXO) and the Very Large Array (VLA), which allowed the discovery of radio and X-ray afterglows.  The XMM Observatory reported an X-ray flux in the 2- 10 keV energy range of $(3.95\pm 0.09)\times 10^{-13}\,{\rm erg\, cm^{-2}\,s^{-1}}$. The CXO reported an X-ray count rate in the 2- 10 keV energy range of $5.6\times 10^{-4}\,{\rm s^{-1}}$.  The redshift of GRB 031203 was determined to be $z=1.055\pm0.0001$ \citep{2004ApJ...611..200P,2004ApJ...605L.101W,2007A&A...474..815M}.  Spectral analysis revealed an isotropic-equivalent energy in the $1 - 10^4\,{\rm keV}$ energy range of $E_{\rm \gamma, iso}=1.67^{+0.04}_{-0.10}\times 10^{50}\,{\rm erg}$.   At $\approx 10$ hours, the corresponding isotropic X-ray luminosity was $9\times 10^{42}\,{\rm erg}$, almost $10^{3}$ fainter than the usual observed burst. At 8.5 GHz, the peak luminosity corresponded to $\approx 10^{29}\,{\rm erg\,s^{-1}\, Hz^{-1}}$, $10^{2}$ fainter than the usual observed burst.

\subsubsection{GRB 060218}

GRB 060218 was detected by the {\itshape Swift} / BAT instrument on February 18, 2006, at 03:34:30 UT in the coordinates RA(J2000)$=03^{\rm h} 21^{\rm m} 37^{\rm s}$ and Dec(J2000)$=+16^{\circ} 51^{\prime} 58^{\prime \prime}$ \citep{2006GCN..4775....1C}. Afterglow confirmation from the X-ray Telescope (XRT) instrument aboard {\itshape Swift} was a $T+153$~s detecting a bright, fading point source \citep{2006GCN..4776....1K}. Simultaneously, the Ultraviolet and Optical Telescope (UVOT) aboard {\itshape Swift} tracked this event for up to $1036$ s after the initial trigger \citep{2006GCN..4779....1M}. The redshift of the event was determined by its association with a host galaxy at $z=0.0331$ \citep{2006GCN..4792....1M}. This event was associated with SN 2006aj.  The XRT instrument monitored GRB 060218 in the windowed-timing (WT) mode with a spectrum exposure of 2.6~ks and the Photon Counting (PC) mode with a spectrum exposure of 52.8 ks. The best-fit values for the intrinsic absorption column densities were $3.58^{+0.07}_{-0.07}\times 10^{21}\,{\rm cm^{-2}}$ and $4.4^{+0.6}_{-0.6}\times 10^{21}\,{\rm cm^{-2}}$ for the WT and PC modes, respectively.  A black-body component exhibiting a decreasing temperature of 0.17 keV and a simultaneous increase in brightness with time was observed. This black-body was observed at $7\times 10^{3}\,{\rm s}$ with a temperature of $0.10\pm 0.05\,{\rm keV}$. At $1.2\times 10^{5}\,{\rm s}$, the black-body was detected by UVOT with a temperature of $3.7^{+1.9}_{-0.9}\,{\rm eV}$, implying an expansion with a sub-relativistic velocity.

\subsubsection{GRB 100316D}
GRB 10031D was detected by the {\itshape Swift} / BAT instrument on 16 March 2010 12:44:50 UT at RA(J2000)$=07^{\rm h} 10^{\rm m} 28^{\rm s}$ and Dec(J2000)$=-56^{\circ} 16^{\prime} 40^{\prime \prime}$  \citep{2010GCN.10496....1S}. Posterior analysis of the BAT light curve photometry found temporal and spectral similarities with SN GRB 060218/2006aj \citep{2010GCN.10511....1S}. The identification of the redshift was done using observations carried out by the VLT Collaboration \cite{2010GCN.10512....1V}. Using data from the X-shooter spectrograph, \citep{2010arXiv1004.2262C} reported the association of GRB 10031D with a galaxy located at $z=0.059$. This event was associated with SN 2010bh. The XRT instrument monitored this burst in the WT mode with a spectrum exposure of 594~s and the PC mode with a spectrum exposure of 3.8~ks. The best-fit values for the intrinsic absorption column densities were $1.06^{+0.06}_{-0.06}\times 10^{22}\,{\rm cm^{-2}}$ and $5.6^{+17.9}_{-5.5}\times 10^{21}\,{\rm cm^{-2}}$ for the WT and PC modes, respectively.  This burst was also detected by Gemini-South and HST for almost two weeks.  A black body component contributing only $\sim 3\%$ of the X-ray flux in the 0.3 - 10 energy range was detected with a temperature of $0.14\pm 0.01\,{\rm keV}$ and a luminosity of $3-4\times 10^{46}\,{\rm erg\,s^{-1}}$.

\subsubsection{GRB 130603B}

On 3 June 2013 at 15:49:14, GRB 130603B was simultaneously detected by the BAT instrument aboard the \textit{ Swift} satellite \citep{2013GCN.14735....1M} and by Konus-Wind \citep{2013Natur.500..547T}. Its location was found to be at $\textrm{R.A.}=+21^{\textrm{h}} 23^{\textrm{m}}27^{\textrm{s}}$, $\textrm{Dec}=-53^{\circ}02'02''$ (J2000). According to the BAT instrument, it had a duration of $T_{90}\approx0.18\pm0.02\,{\rm s}$ in the 15-350 keV band \citep{2013GCN.14741....1B}, which places it in the sGRB class. The {\itshape Swift}/XRT instrument began to observe the error box at $T+43$~s discovering an uncatalogued X-ray source at the location of the burst \citep{2013GCN.14739....1E,2013GCN.14749....1K}. Subsequent X-ray observations were performed days later by the \textit{XMM-Newton} telescope \citep{2013GCN.14922....1F}. Using ground telescopes, this burst was followed-up by the twin \textit{Magellan} telescopes in the optical band \citep{2013GCN.14745....1F}, whereas in radio by the VLA (Very Large Array) \citep{2013GCN.14751....1F}. Optical observations allowed the determination of the redshift of $z=0.3568\pm0.0005$ \citep{2013ApJ...777...94C}. Optical and near-IR observations of the event also demonstrated the presence of excess near-IR emission matching a KN \citep{2013ApJ...774L..23B}.

\subsubsection{GRB 150101B}

The {\itshape Swift}/BAT and the {\itshape Fermi}/GBM instruments triggered GRB 150101B on January 1, 2015, at 15:23 and at 15:24:34.468, respectively \citep{2015ATel.6871....1C,2018ApJ...863L..34B} with an estimated duration of about $T_{90} = 0.08\pm0.93$~ s. Features of absorption in the spectrum at lower frequencies were carried out with {\itshape Magellan}/Baade, \textit{VLT} and \textit{TNG} \citep{2016ApJ...833..151F}, associated with a young host galaxy located in the constellation Virgo at $z=0.1343\pm0.0030$.  The total fluence and isotropic-equivalent gamma-ray energy in the energy range of 10 - 1000 keV were $\sim 10^{-7}\,{\rm erg\, cm^{-2}}$ and $E_{\rm \gamma,iso}\sim 4.7\times 10^{48}\,{\rm erg}$, respectively \citep{2018NatCo...9.4089T}, which  corresponded to one of the lowest energies reported by Swift/BAT.  The characteristics of this burst were very similar to those of the gravitational wave event GW170817, associated with a merger of a binary NS stars \citep{2018ApJ...863L..34B, 2018NatCo...9.4089T}.

\subsubsection{GRB 171205A}

The BAT instrument triggered and localized GRB 171205A on 2017 December 5 at 07:20:43.9 UT \citep{GCN22177}. With an isotropic luminosity $L_{\rm iso,\gamma}=3.0\times 10^{47}\,{\rm erg\,s^{-1}}$ and a duration of $T_{90}=190.0\pm 35.0\,{\rm s}$ in the 15-150-keV energy range \citep{2017GCN.22184....1B}, GRB 171205A was classified as a long llGRB~\citep{2019Natur.565..324I}.  This GRB was also detected by Konus-\textit{Wind} \citep{GCN22227} and was observed by Swift-XRT approximately 134 s after the BAT trigger \citep{GCN22183}. Swift/UVOT confirmed a source consistent with the XRT position, so GRB 171205A was localized with coordinates RA(J2000)=11:09:39.547 and Dec(J2000)=-12:35:17.93 \citep{GCN22181}. Several ground-based experiments were able to start observing the GRB in the optical and infrarred bands only a few minutes after the initial BAT trigger, such as the CIBO collaboration \citep{GCN22189}, the Gao-Mei-Gu station \citep{GCN22186} and SNUCAM-II \citep{GCN22188}, among others. Five days later, the 10.4-m Gran Telescopio Canarias (GTC) picked up on the emergence of the associated type Ic supernova \citep[SN 2017iuk;][]{2017ATel11038....1D} and SMARTS 1.3-m telescope \citep{2017GCN.22192....1C}. These detections were able to confirm the previous association between the burst and its host galaxy 2MASX J11093966-1235116 by \cite{GCN22178} and to also determine its redshift through the detection of absorption and emission lines at $z = 0.0368$ \citep{GCN22180,2018A&A...619A..66D}. GRB 171205A has been widely studied since it was the third nearest burst.\\ 

Since GRB 171205A was so close, \cite{2019Natur.565..324I} followed-up multi-wavelength observations using photometry and spectroscopy.  They presented spectroscopic observations of the supernova SN 2017iuk taken across many time periods.  Features at extraordinarily high expansion velocities ($\sim 1.15\times 10^5\, {\rm km\, s^{-1}}$) were seen in the spectra during the first day following the burst.  They used spectral synthesis models created for SN 2017iuk to demonstrate that the chemical abundances of these characteristics were distinct from those seen in the ejecta of SN 2017iuk at subsequent dates. The authors went on to demonstrate that a hot cocoon produced by the expansion and deceleration of an ultra-relativistic jet inside the progenitor star into the circumstellar medium was where the high-velocity characteristics are originating. During the first three days after the explosion, the cocoon material becomes transparent and dominates the electromagnetic emission and, later, the supernova emission begins to dominate the emission, and this cocoon quickly becomes opaque and is overtaken by the supernova emission.\\

\subsection{Results and Discussion}

Figures~\ref{lc_GRB980425}, \ref{lc_GRB031203}, \ref{lc_GRB060218}, \ref{lc_GRB100316D}, 
\ref{GRBs_130603B-150101B_lc_ISM}, 
\ref{GRBs_171205_lc_wind} and  \ref{GRBs_171205_lc_ISM} show the multiwavelength observations of GRB 980425, 031203, 060218, 100316D, 130603B, 150101B and  171205A, respectively, with the best-fit curves generated with our analytical scenario.   We have considered the synchrotron radiation from the quasi-spherical material that decelerates in the stellar-wind environment (left panel) and ISM (right panel) for GRB 980425, 031203, 060218, 100316D and 171205A, and in the cases of the short bursts  GRB 130603B and  150101B we consider the deceleration in ISM.  We have implemented Markov-Chain Monte Carlo (MCMC) simulations using eight (E, $A_{\rm st}$, $\Gamma_0$, $\varepsilon_{\rm B}$, $\varepsilon_{\rm e}$, $p$, $\alpha_{\rm s}$, $\theta_c$) parameters from the entire sample of llGRBs to determine the optimal values that characterize the multiwavelength observations with our synchrotron model.   To describe the entire data in our sample, a total of 17,200 samples and 5,100 tuning steps were used.   We only show Figure~\ref{mcmc_GRB980425}, which corresponds to GRB 980425, for displaying  the best-fit parameter values with the respective median of the posterior distributions. In Table~\ref{best_fit_par}, the best-fit values are shown in red, and the median of the posterior distributions is presented.

The optimal values of the magnetic microphysical parameters fall within the range of $10^{-2}\lesssim \varepsilon_{\rm B}\lesssim 10^{-1}$. The best-fit values of the  microphysical parameter $\varepsilon_{\rm e}$ are within the range of $0.93\times10^{-2}\lesssim\varepsilon_{\rm e}\lesssim0.80$.  Considering the magnetic equipartition parameter, the magnetization parameter is predicted to fall within a certain range. Taking into account the $\varepsilon-$values, the magnetization parameter is expected to be between $10^{-2}$ and $10^{-1}$, which means that the material are moderately magnetic. There has been previous work with the intent to model afterglow observations of those GRBs we have considered in this work \citep{2006Natur.442.1008C, 2007A&A...465....1D, 2011ApJ...726...32F, 2013ApJ...778...18M, 2016MNRAS.460.1680I, 2019Natur.565..324I, 2021ApJ...907...60M}. The values we have obtained for both microphysical parameters are in good agreement with the aforementioned studies.   \cite{2014ApJ...785...29S} performed a statistical and bibliographical study to determine the distribution of $\varepsilon_{\rm B}$ in afterglow models. Their results showed that in both the optical and X-ray bands the distribution is rather wide, with optical measurements favoring $\varepsilon_{\rm B}\sim10^{-9}-10^{-3}$ while X-ray observations prefer $\varepsilon_{\rm B}\sim10^{-5}-10^{-1}$. The optimal values that we have obtained in our work are well in agreement with the later observations. In order to explain such large values of $\varepsilon_{\rm B}$, \cite{2014ApJ...785...29S} addressed the possibility of bursts taking place in environments with particularly high seed
magnetic fields ($B_0$).



The most suitable values of half-opening angles, lie in a range of $32^\circ\lesssim\theta_c\lesssim 55^\circ$, which are completely different from the values reported for half-jet opening angle of classical GRBs  $\theta_j=2^\circ-10^\circ$ \citep{2001ApJ...562L..55F,2015ApJ...815..102F,2005ApJ...620..355L,2013ApJ...777..162M,2018ApJ...857..128J}.  The large values found with our MCMC simulations agree with the multiwavelength observations which show no indication of late, steep decays over a time scale of weeks.  It is worth noting that in some bursts show no indication of late steep decays in a time scale of days. For instance, GRB 190829A,  located at $z = 0.0785 \pm 0.005$ \citep{2019GCN.25565....1V} and classified as an intermediate-luminosity burst $E_{\rm \gamma, iso}=(2.967\pm 0.0032)\times 10^{50}\,{\rm erg}$, was one of the closest bursts, which was detected in TeV gamma- and X-rays by H.E.S.S. and Swift-XRT.  The H.E.S.S. and Swift-XRT observations did not exhibit a jet break up to $\sim 10^6\,{\rm s}$, indicating that the half-jet opening angle was larger than $ \gtrsim 17^\circ$ \citep[e.g., see][]{2021ApJ...918...12F, 2021arXiv210603466L, 2021MNRAS.504.5647S}.

The best-fit values of the power-law indexes for the electron distribution and the velocity distribution are in the ranges of $2.1\lesssim p \lesssim 3.5$ and  $2\lesssim \alpha_{\rm s}\lesssim 3$, respectively, which is in the range of the ones reported in literature \citep{1998Natur.395..670G, 2006Natur.442.1008C, 2007A&A...465....1D, 2011ApJ...726...32F, 2013ApJ...778...18M, 2016MNRAS.460.1680I, 2019Natur.565..324I, 2021ApJ...907...60M}.  For instance, the late-time radio and X-ray observations and spectral/temporal analysis of GRB 100316D were modelled by \cite{2013ApJ...778...18M}, reporting a value close to  $\alpha_{\rm s}\sim 2.4$, and associating the progenitor as the emergence of the fastest rotating magnetar.

Efficiency is essential for understanding the gamma-ray emission mechanism. Considering the optimal values of the equivalent kinetic energies are $10^{50}\lesssim E\lesssim 10^{52}\,{\rm erg}$ and the isotropic energies in gamma-rays during the early phase in the range of $10^{48}\lesssim E_{\rm \gamma, iso}\lesssim 10^{50}\,{\rm erg}$, kinetic efficiencies fall around $\eta \approx 1 - 5\%$, which are typical compared to those values reported \citep{2006Natur.442.1008C, 2007A&A...465....1D, 2011ApJ...726...32F, 2013ApJ...778...18M, 2016MNRAS.460.1680I, 2019Natur.565..324I, 2021ApJ...907...60M}.

The best-fit values of the density lie in the range of $ 0.1\lesssim A_{\rm st} \lesssim 10$. \cite{2015MNRAS.448..417B} have considered the radio light curves of the late afterglow of GRB 980425, 031203, 060218 and 100316D. The initial bulk Lorentz factor values that are best-fitted fall within the $1.7\lesssim \Gamma_0\lesssim 4.9$, which is similar to the values found in other bursts.  In the case of GRB 9804125, they used the non-relativistic phase description for an ISM medium, however the analytic behavior at late times increased with time, in contrast with observations. Therefore, they considered a wind medium with $A_{\rm st}=10$ and $p=2.1$ which was able to reproduce the radio and X-ray observations. Their values were in agreement with \cite{1999ApJ...526..716L}, who previously modeled the radio data of GRB 980425. They made the case for a surrounding wind medium. They presented two possible solutions, one with $A_{\rm st}\approx6$ for $\varepsilon_{\rm e}\sim 1$ and $\varepsilon_{\rm B}=10^{-6}$, and another with $A_{\rm st}\approx0.04$ with $\varepsilon_{\rm e}=\varepsilon_{\rm B}=0.1$.  In the case of GRB 031203, \cite{2015MNRAS.448..417B} and \cite{2004Natur.430..648S} preferred an ISM model with $n\approx0.1\ \mathrm{cm}^{-3}$ due to the rather flat behavior of optical observations at late times.    Regarding GRB 060218 and 100316D, \cite{2015MNRAS.448..417B} found good agreement in the radio bands for a wind medium with $A_{\rm st}=1$ and $\varepsilon_{\rm B}\sim10^{-5}$, but their model predicted an X-ray flux lower than observations in both bursts.   In the case of GRB 171205A, \cite{2021ApJ...907...60M} made the case for a stratified wind-like medium. They worked with both the standard isotropic afterglow model and a shock breakout model, for which they found that values of $A_{\rm st}$ in the range $\sim0.1-1$ were able to reproduce radio measurements over 1000 days. Spectral and temporal analysis of the radio observations led to the conclusion that the ejecta was expanded to relativistic velocities \citep{1998Natur.395..670G}.  \cite{2007ApJ...667..351W} and \cite{2013ApJ...778...18M} modelled GRB 060218 and GRB 100316D the multiwavelength observations with a mildly relativistic material ($\Gamma=1.5-2$), respectively. 

Our model builds upon the work of~\cite{2015MNRAS.448..417B}, who successfully explained the radio afterglows of GRBs 980425, 060218, and 100316D using a shock-breakout scenario but consistently underpredicted the observed X-ray emission. By employing a quasi-spherical outflow decelerated in a stratified environment, and incorporating a velocity-structured ejecta, self-absorption, and coasting-to-deceleration transitions, our model provides an improved multiwavelength fit. In particular, while we replicate radio observations well and better capture early X-ray features, we remain with problems in the late-time X-ray emission, especially for GRB 980425. The flux density remains underpredicted, similar to~\cite{2015MNRAS.448..417B}, for both constant and wind-like media, with the constant-medium scenario being the most preferable.

\cite{2013ApJ...778...18M} considered of the kinetic equivalent energy ($10^{46}\lesssim E \lesssim 10^{53}\,{\rm erg}$) and the profile of velocity ($ 10^{-2}\lesssim \Gamma\beta \lesssim 30$) distribution of the ejected material from ordinary Type Ib/c \citep{2003ApJ...599..408B,2006Natur.442.1014S,2008Natur.453..469S, 2010ApJ...725..922S, 2010Natur.463..513S, 2012ApJ...756..184S, 2013MNRAS.434.1098C, 2013ApJ...770L..38M, 2013MNRAS.432.2463M, 2013arXiv1309.3573K}, from sub-energetic bursts \citep{2003Natur.426..154B, 2006ApJ...638..930S, 2013MNRAS.434.1098C}, from Relativistic SN 2009bb \citep{2010Natur.463..513S, 2013MNRAS.434.1098C}, and from classical bursts \citep{2003ApJ...599..408B, 2006ApJ...646L..99F, 2008ApJ...683..924C,  2013MNRAS.434.1098C, 2015ApJ...804..105F, 2016ApJ...818..190F, 2017ApJ...848...94F}. Our sample of LL GRBs belongs to the sub-energetic bursts with similar values.


\cite{2008ApJ...672..433S} analysed a sample of 10 GRBs. They showed that four sources were consistent with a wind-like medium, while three were not. However, these last three were not necessarily well-fit by a constant medium, but rather with a stratification index $0\leq k\leq1$. \cite{2009MNRAS.395..580C} constrained parameters for 10 GRBs by analysing their optical and X-ray light curves. They found that two of their GRBs were consistent with both $k=0$ and $k=2$, two only with $k=0$ and two only with $k=2$. Therefore, they concluded that the circumburst environments may be drawn from either of the two possibilities and they also discussed the possibility that $k\neq 0$ or $2$. \cite{2011AIPC.1358..165S} assembled a sample of 27 Swift-detected GRBs (including one short GRB) and made observations about the density profiles of their environments using the observed X-ray and optical afterglow emission. They determined that out of their sample, 18 GRBs are consistent with environments that are homogeneous, while 6 GRBs are consistent with wind-like media. \cite{2022MNRAS.511.2848A} used Bayesian inference to infer the best-fit parameters for a sample of 22 long GRBs and 4 short GRBs. They considered both a $k=0$ and a $k=2$ model and found that there is approximately an even split between homogeneous and wind-like environments for the case of long GRBs.

Based on the jet break observed in the radio and optical bands of GRB 130603B,  \cite{2014ApJ...780..118F} suggested a transition in the afterglow emission.  The authors reported the best-fit value of the inferred jet opening angle was in the range of $\approx 4^\circ - 8^\circ$, which corresponded to a beaming-corrected kinetic energy in the range of $E\approx (0.1 - 1.6)\times 10^{49}\,{\rm erg}$.  \cite{2014ApJ...780..118F} also reported a constant-density value of $n=5\times10^{-3}-30\ \mathrm{cm}^{-3}$ from uncertainty of the self-absorption frequency in the standard synchrotron model.    On the other hand, \cite{2015MNRAS.450.1430H} were able to fit the radio light curves with $n=1.0\ \mathrm{cm}^{-3}$ and the microphysical parameters $\varepsilon_{\rm e}=0.2,\,\varepsilon_{\rm B}=8\times10^{-3}$. However, the possible range of the circumburst density spans several orders of magnitude, so with this in mind, \cite{2019MNRAS.485.5294P} were able to fit light curves from microwaves to X-rays with the parameter values of $n\approx0.004\ \mathrm{cm}^{-3}$, $\varepsilon_{\rm e}\approx 0.07,\,\varepsilon_{\rm B}\approx 0.03$.  For GRB 150101B, 
\cite{2016ApJ...833..151F} used the X-ray afterglow observations to constrain the circumburst density to $(0.8-4)\times10^{-5}\ \mathrm{cm}^{-3}$ for $\varepsilon_{\rm B}=0.01-0.1$. More recently, \cite{2020ApJ...896...25F} were able to fit the X-ray light curve with an ISM model with $n\approx10^{-2}\ \mathrm{cm}^{-3}$ for the microphysical parameters $\varepsilon_{\rm B}\approx10^{-2},\,\varepsilon_{\rm e}\approx10^{-1}$, and with a stellar wind model with $A_{\rm st}\approx10^{-1}$ for $\varepsilon_{\rm B}\approx6\times10^{-2},\,\varepsilon_{\rm e}\approx0.9\times10^{-1}$.\\

\cite{2014ApJ...780..118F} modeled the afterglow of short GRB 130603B using a standard synchrotron scenario from a collimated jet in a uniform medium. While they achieved good fits to the radio and optical data, their model significantly underpredicted the late-time X-ray emission. In contrast, our quasi-spherical outflow model, which accounts for stratified environments, velocity-structured ejecta, and self-absorption, successfully reproduces the multiwavelength observations, including the X-ray excess.

\subsection{A burst sample reported in Swift satellite data base}

It was unexpected to discover a GW event GW170817 at a close distance of $\approx$ 40 Mpc \citep{PhysRevLett.119.161101, 2041-8205-848-2-L12}. A limited population of feeble gamma-ray transients following NS mergers was uncovered.  GW170817 was linked to the faint and early gamma-ray emission of GRB 170817A \citep{2017ApJ...848L..14G, 2017ApJ...848L..15S} and the kilonova AT2017gfo \citep{2017Sci...358.1556C}, which had an initial intense optical emission with an absolute magnitude of around $-16\,{\rm mag}$, 12 hours after the NS merger. This was followed by a very late multiwavelength afterglow, reaching its peak at $\approx$ 150 days after this merger \citep{troja2017a, 2017Sci...358.1559K, 2018MNRAS.478..733L, 2018ApJ...867...57R, 2017ApJ...848L..20M, 2017ApJ...848L...6L, 2018MNRAS.479..588G, 2019ApJ...884...71F, 2018MNRAS.479..588G, 2019ApJ...871..200F}.\\

\cite{2020MNRAS.492.5011D} conducted a methodical search in the Swift database spanning 14 years to find neighboring sGRBs like GRB 170817A. Four possible prospects, GRB 050906, GRB 070810B, GRB 080121, and GRB 100216A, were discovered within a range of 100 to 200 ${\rm Mpc}$. They employed these potential sources to restrict the features of X-ray counterparts of the merging of two neutron stars, and calculated optical upper bounds for the beginning of a "blue" kilonova, suggesting a small quantity of lanthanide-deficient material.

\subsubsection{Multi-band observations}
We present the multi-band observations of the closest short bursts detected by Swift satellite.

\paragraph{GRB 050906}  GRB 050906 triggered the Burst Alert Telescope (BAT) on board the Swift satellite at 2005 September 5 10:32:18 UTC with a reported location of $\textrm{R.A.}=03^{\textrm{h}}31^{\textrm{m}}13^{\textrm{s}}$, $\textrm{Dec}=-14^{\circ}37'30''$ (J2000) with a positional accuracy of $3'$\citep{GCN3926}.
The light curve showed an excess in the 25-100 keV energy range. The duration and the observed fluence in the energy range of 15 - 150 keV were $128\pm16\ \textrm{ms}$ and $(5.9\pm3.2)\times10^{-8}\ \textrm{erg}\ \textrm{cm}^{-2}$, respectively \citep{GCN3935}. Details of the deep optical and infrared observations collected with Swift are reported in \cite{2008MNRAS.384...541D}. According to the authors, neither X-ray nor optical/IR afterglow was discovered to deep limits.

\paragraph{GRB 070810B}  GRB 070810B triggered Swift BAT at 2007 August 10 15:19:17 UTC with a reported location of $\textrm{R.A.}=00^{\textrm{h}}35^{\textrm{m}}46^{\textrm{s}}$, $\textrm{Dec}=+08^{\circ}50'07''$ (J2000) with a positional accuracy of $3'$\citep{GCN6743}. Details of the follow-up observations carried out by the KANATA 1.5-m telescope, the Xinlong TNT 80 cm telescope, the 2-m Faulkes Telescope South, the Shajn 2.6 m telescope and the Keck I telescope (HST) are summarized in \cite{2019Preprint}. Only the Shajn telescope detected a source inside the error box of GRB 070810B \citep{GCN6762}.

\paragraph{GRB 080121}  GRB 080121 triggered Swift BAT at 2008 January 21 21:29:55 UTC with a reported location of $\textrm{R.A.}=09^{\textrm{h}}08^{\textrm{m}}56^{\textrm{s}}$, $\textrm{Dec}=+41^{\circ}50'29''$ (J2000) with a positional accuracy of $3'$. The duration and the observed fluence in the 15-150 keV energy range were $T_{90}=0.7\pm0.2\ \textrm{s}$ and $(3\pm2)\times10^{-8}\ \textrm{erg}\ \textrm{cm}^{-2}\ \textrm{s}^{-1}$, respectively \citep{GCN7209}. Follow-up observation were carried out 2.3 days after the burst using UVOT and XRT instruments, but neither X-ray afterglow candidate nor sources were found within the BAT error circle \citep{GCN7217,GCN7224}. Two galaxies were present within the BAT error circle, which would associate a redshift of $z\sim0.046$ to GRB 080121, but the isotropic energy released would be several orders of magnitude less than in typical short-hard bursts \citep{GCN7210}.

\paragraph{GRB 100216A} GRB 100216A triggered Swift BAT and Gamma-ray Burst Monitor (GBM) aboard the Fermi spacecraft at 2010 February 16 10:07:00 UTC with a reported location of $\textrm{R.A.}=10^{\textrm{h}}17^{\textrm{m}}03.2^{\textrm{s}}$, $\textrm{Dec}=+35^{\circ}31'27.5''$ (J2000) with a positional accuracy of $3'$. The duration and the observed fluence of the single peak in the energy range of 15 - 350 keV were $0.3\ \textrm{s}$ and $(4.7\pm3)\times10^{-8}\ \textrm{erg}\ \textrm{cm}^{-2}$, respectively \citep{GCN10428}. Follow-up observations were conducted by XRT and UVOT from $214.4\ \textrm{ks}$ to $249.2\ \textrm{ks}$ after the BAT trigger. No fading source was detected within the observation, but a source thought to be 1RXS J101702.9+353404 was identified within the error circle \citep{GCN10435,GCN10442}.

\subsubsection{Analysis and Description}

Figure~\ref{fig:SSGRBs} presents the parameter space of rejected values of ($\varepsilon_B,\varepsilon_e,\beta$) for GRB 050906, 080121 and 100216A. The other parameters of the model have been fixed to $n=1\,\mathrm{cm}^{-3}$, $E=10^{50}~\rm erg$, $\alpha_s=2$ and $p=2.2$ for all GRBs under consideration. Here, we define rejection as a set of parameters for which the flux overcomes the observational upper limits. Regions without color represent that any value of $\beta$ is allowed and remains below the upper limits. On the other hand, areas of a particular color mean that any values larger than the corresponding $\beta$ are rejected.   The constraints are the most stringent in the GRB 050906 (left panel). A broad region of the $(\varepsilon_e, \varepsilon_B)$ space is colored, indicating that for a wide range of microphysical parameters, only sub-relativistic low values of $\beta$ are compatible with the observational limits, namely $\beta\lesssim0.7$. Contrastingly, GRB 080121 (middle panel) shows the least constrained parameter space. The uncolored region dominates the plot, which means that for most values of $\varepsilon_e$ and $\varepsilon_B$, the modeled flux remains below the upper limits of the observation regardless of $\beta$. Only the top right corner, which corresponds to the most significant possible microphysical parameters, showcases some limits. Still, the constraints are relatively weak with $\beta\lesssim0.85$, which still agrees with a relativistic shock. Finally, GRB 100216A (right panel) presents a relatively unconstrained scenario. The colored region is also limited to the top right corner, suggesting that while some parameter combinations are excluded for higher $\beta$, the constraints are generally weaker than in GRB 050906. Compared to GRB 080121, smaller values of $\beta$ are required. The panel associated with GRB 070810's parameter space is not shown because the upper limits do not constrain any parameter region. 

\section{New points in our approach and previous studies}\label{sec5}

The quasi-spherical materials in the short- and long-burst scenarios have been associated with the cocoon  \citep{2014ApJ...784L..28N, 2014ApJ...788L...8M,2017ApJ...848L...6L, 2018PhRvL.120x1103L, 2017ApJ...834...28N,2018MNRAS.473..576G} and shock breakout materials  \citep{1994ApJ...431..742D, 2006ApJ...645..431N, 2009ApJ...690.1681D, 2011MNRAS.414.1715B, 2013ApJ...778...18M, 2014ApJ...784L..28N, 2014MNRAS.437L...6K, 2015MNRAS.448..417B, 2015MNRAS.446.1115M, 2018NatAs...2..808F, 2019ApJ...871..200F, 2021ApJ...910..128H, 2022ApJ...933..164G}.   \cite{2014ApJ...784L..28N} and \cite{2014ApJ...788L...8M} derived the parameter space of Lorentz factor, duration and kinetic luminosity for short GRB production.  \cite{2014ApJ...788L...8M} concluded  that if the jet successfully penetrates the neutrino-driven wind, the energy contained in the cocoon may enhance the precursor. They found that the cocoon material could be ejected with Lorentz factors in the $\Gamma\sim2 - 3 $  range.  \cite{2017ApJ...834...28N} explored cocoon scenarios with Lorentz factors around $\sim 10$.   The shock breakout is traditionally expected to be a brief event lasting hours. However, \cite{2018NatAs...2..808F} analyzed observations of 26 supernovae and presented evidence that the shock breakout is significantly delayed and extended due to dense circumstellar material surrounding the star in most SNe II. This material causes the shock to emerge gradually over several days, rather than hours.   Similarly, \cite{2021ApJ...910..128H} also explored how dense circumstellar material around a star influences the shock breakout during a supernova event. The authors also concluded that the interaction with a dense circumstellar material can extend the shock breakout's duration and increase its brightness. They focused on the case of the supernova PS1-13arp, which exhibited an early UV excess indicative of such an interaction.   \cite{2022ApJ...933..164G}  utilized Athena++ to conduct 3D Radiation-Hydrodynamic simulations of the radiative breakout of a shock wave in the outermost layer of a red supergiant (RSG) that has undergone core collapse and is expected to evolve into a Type IIP supernova. \cite{2011MNRAS.414.1715B} proposed a simple scenario to estimate the conditions during shock breakout in a stellar wind. This model was based on the observable parameters in the X-ray flash light curve, such as the total X-ray energy and the diffusion time after the maximum flux, and provided relationships for the radius and velocity of the shock at the moment of breakout. The derivation required the self-similar solution for the structure of a forward-reverse shock that is anticipated when ejecta moves through a pre-existing wind at significant distances from the surface of the progenitor. \cite{2013ApJ...778...18M} conducted a spectral and temporal study of GRB 100316D using late-time radio and X-ray data. The results showed a mildly relativistic material moving with $\Gamma=1.5-2$, and $\alpha_{\rm s}\sim 2.4$, indicating that the progenitor of the GRB was likely the formation of a magnetar with extremely rapid rotation.  \cite{2014MNRAS.437L...6K} considered an ultrarelativistic shock breakout ejected from a binary NS and decelerated by a constant circumstellar medium. They estimated the  radio, optical, and X-ray fluxes in timescales from seconds to days after the trigger time.  \cite{2015MNRAS.448..417B} proposed that the prompt and afterglow emission of some long llGRBs (GRB 980425, 031203, 060218, 100316D) could be described by the deceleration of the relativistic shock breakout material in a stellar wind and ISM. The authors observed differences in X-ray fluxes, suggesting unconsidered environmental complexity. \cite{2015MNRAS.446.1115M} and  \cite{2015MNRAS.450.1430H} proposed a shock breakout material from a binary NS and estimated the radio fluxes on timescale of days to months.  \cite{2019ApJ...871..200F} derived the synchrotron light curves of a shock-breakout material when it decelerates in a uniform medium and also applied this phenomenological model to describe the multiwavelength afterglow emission up to a pair of weeks of GRB 170817A.\\ 

Previous studies of GRB afterglows often assume a circumburst environment modeled as a constant density medium ($k=0$) or a stellar wind ($k=2$), with fixed microphysical parameters and a standard electron spectral index ($p>2$).  These assumptions face challenges in explaining the stratified density profiles inferred from observations of GRB-SNe systems \citep[e.g., see][]{2013ApJ...774L..23B, 2019Natur.565..324I, 2020ApJ...905..112F}. In some cases, such as GRBs 980425 and 060218, the multiwavelength observations have been partially described, and more than one electron population is required. For instance, \cite{2015MNRAS.448..417B} considered a wind shock-breakout model to fit the radio data of GRB 980425. However, the same afterglow model used to fit the radio led to a very low X-ray flux and soft X-ray spectrum, contrasting with the observations. However, the authors showed that the model can simultaneously explain the radio and X-ray observations with a modification introducing two power-law electron distributions; a distribution with $p=2.8$ for the radio-emitting electrons and $p=2.1$ for the X-ray-emitting ones. \cite{2015MNRAS.448..417B} explained the prompt emission of GRB 060218 as arising from a mildly relativistic shock breakout, and its radio afterglow from the interaction between ejecta and the surrounding wind medium. They found that the shock remained in a coasting phase, with a Lorentz factor close to $1.3$, until deceleration occurred around $200$ days. While they could fit the radio data with this model, the predicted X-ray afterglow significantly underpredicted the observed X-ray flux. The authors suggested that the X-ray and radio emissions might originate from different electron populations or emission processes. The authors found good agreement in the radio bands for a wind medium with $A_{\rm st}=1$, $\varepsilon_{\rm B}=1.5\times 10^{-5}$, $\varepsilon_{\rm e}=0.2$, and $p=2.6$.  To overcome this limitation, we have systematically generalized the afterglow model to $0\leq k<3$, allowing for analyzing environments beyond ISM or wind, such as those predicted for merger ejecta or progenitors with non-steady mass loss \citep{2005ApJ...631..435R, 2012ApJ...747..118M}.
Additionally, we have broadened the synchrotron spectra to encompass the amount of electrons accelerated by the shock front and the self-absorption regime, which is essential for modeling early radio emission \citep{2019ApJ...884L..58U}. We have also unified the relativistic and sub-relativistic regimes, facilitating continuous modeling from prompt cocoon or breakout material to late-time emission.

\section{Conclusion}\label{sec4}

We have extended the analytical synchrotron afterglow scenario of the quasi-spherical material previously utilized to characterize the early X-ray, optical and radio afterglow observations from GRB 170817A \citep{2019ApJ...884...71F}. Binary compact objects are predicted to have a uniform-density medium ($k=0$), but massive stars with varying mass-loss evolutions ($\dot{M}$ or $v_{\rm w}$) towards the end of their lifetimes are expected to consist of a stratified medium \citep[$0\leq k < 3$;][]{2005ApJ...631..435R,  2006A&A...460..105V}. In the current model, we have generalized and derived the light curves to a stratified environment with a density profile $ \propto r^{-k}$ with $0\leq { k}< 3 $, including the coasting phase, the self-absorption regime and  considering the possibility that not all electrons are accelerated by the shock front, but just a fraction of them. 

We calculated synchrotron light curves for both fast- and slow-cooling scenarios during the coasting and deceleration phases. In the coasting phase, we considered velocities between $0.07<\beta<0.8$, and for the deceleration phase, we used a power-law velocity distribution $\propto\beta^{-\alpha_{\rm s}}$ with $3<\alpha_{\rm s}<5.2$ for a source located at 100 Mpc. For typical parameter values of GRB afterglows, we presented synchrotron light curves for radio at 6 GHz, optical at 1 eV, and X-rays at 1 keV. The majority of light curves peak in a matter of months or years. On the other hand, the light curve peak might happen in a matter of hours, days or weeks if the quasi-spherical outflow is extremely energetic or slows down in a dense medium.  We showed that density parameter variations are more easily observed (i) in radio rather than X-ray light curves, (ii) in stratified rather than uniform-density media, and (iii) for higher values of $\alpha_{\rm s}$. Hence, the transition from a stellar-wind to a uniform-density medium is more noticeable in radio bands.

We showed that a light curve flattening or rebrightening occurs when nonrelativistic material decelerates in a uniform-density medium; rebrightening is less noticeable when decelerating in a stratified medium. Therefore, deceleration of nonrelativistic material from a binary compact object merger is indicated by a flattening or rebrightening over months to years in conjunction with GW detection. On the other hand, a steadily declining flux over several months to years might signify a slowdown of ejecta from the death of a massive star with varying mass loss at the end of its life. We calculated the expected gamma-ray, X-ray, optical, and radio fluxes from synchrotron emission of electrons in forward shocks. These electromagnetic signatures over different timescales and frequencies resemble those around SNe in a uniform-density medium and coincide with GW detections.

We have applied this analytical model to interpret a sample of llGRBs  (GRB 980425, 031203, 060218, 100316D, 130603B, 150101B and 171205A), the Swift-BAT database with sGRBs located between 100 and 200 Mpc (GRB 050906, 070810B, 080121, 080121 and 100216A).  We have used the multiwavelength observations and upper limits to model and constrain the parameters ($\tilde{E}$, $A_{\rm st}$, $\Gamma_0$, $\varepsilon_{\rm B}$, $\varepsilon_{\rm e}$, $p$, $\alpha_s$, $\theta_c$) in our analytical afterglow model through MCMC simulations. As expected, we have found that all of the results are consistent with sub-energetic GRBs.  The large values of half-opening angles found with our MCMC simulations agree with the multiwavelength observations which show no indication of late, steep decays over a time scale of weeks. We propose that some GRBs with no indication of late steep decays in a time scale of days could be associated with large half-opening angles.   


The Fermi GBM initially observed GRB 170817A on August 17, 2017 at 12:41:06.   Just before the GBM trigger, the LIGO Scientific Collaboration and the Virgo Collaboration announced the detection of a GW candidate (GW170817) that matched the position of GRB 170817A \citep{2017GCN.21506....1C,
PhysRevLett.119.161101, 2041-8205-848-2-L12}. The progenitor of GRB 170817A was immediately linked to the merging of two neutron stars, marking the initial observation of GWs \citep{PhysRevLett.119.161101, 2041-8205-848-2-L12}.  GRB 170817A was promptly followed by a comprehensive observing effort that included radio, optical, and X-ray wavelengths \citep[e.g., see][and references therein]{troja2017a, 2041-8205-848-2-L12}.  The non-thermal spectrum collected over the first approximately 900 days following this merger was modelled by synchrotron radiation emitted by an off-axis structure jet \citep[e.g., see][]{troja2017a, 2017Sci...358.1559K,  2018MNRAS.478..733L, 2018ApJ...867...57R, 2017ApJ...848L..20M, 2017ApJ...848L...6L, 2018MNRAS.479..588G, 2019ApJ...884...71F, 2019ApJ...871..200F}.   \cite{2022ApJ...927L..17H} examined the most recent X-ray and radio data of GRB 170817A, which were obtained using CXO, VLA and the MeerKAT around 3.3 years after the merger. \citet{2022ApJ...927L..17H} noticed and reported evidence of an unexpected X-ray emission component, which did not align with the synchrotron off-axis afterglow hypothesis in a constant-density medium.   This X-ray excess would be a candidate to be described with the quasi-spherical outflow model in the sub-relativistic regime.

\section*{Acknowledgements}

We express our sincere gratitude to the anonymous referee for their thorough review of the paper and valuable suggestions that enhanced the clarity of the manuscript. We greatly appreciate the useful discussions with Rodolofo Barniol-Duran and Tanmoy Laskar. NF is grateful to UNAM-DGAPA-PAPIIT for the funding provided by grant IN112525. BBK is supported by IBS under the project code IBS-R018-D3. AG was supported by Universidad Nacional Autónoma de México Postdoctoral Program (POSDOC).

\section*{Data Availability}

No new data were generated or analysed in support of this research.



\bibliographystyle{mnras}
\bibliography{main} 

\addcontentsline{toc}{chapter}{Bibliography}

\clearpage
\newpage

\begin{table}
\centering \renewcommand{\arraystretch}{1.85}\addtolength{\tabcolsep}{1.5pt}
\caption{Evolution of the synchrotron light curves ($F_\nu\propto t^{-\alpha}\nu^{-\beta}$) from a cocoon/shock breakout material decelerated in a stratified environment}
\label{Table2}
\rotatebox{90}{\begin{tabular}{c c c  c c}
\hline \hline
&\hspace{0.5cm}      &\hspace{0.5cm}   Coasting Phase  &Deceleration Phase &\hspace{0.5cm}   Lateral Expansion\\ 
&\hspace{0.5cm}      &Sub-relativ \hspace{0.5cm}   Relativistic     &Sub-relativ \hspace{0.5cm}   Relativistic & \hspace{0.5cm}   Relativistic\\

                     & \hspace{0.5cm}  $\beta $            &   \hspace{0.5cm}  $\alpha $  &  \hspace{0.5cm}  $\alpha $\\  \hline \hline
$\nu_{\rm a,3} < \nu_{\rm c} < \nu_{\rm m} $ \\ \hline

$\nu < \nu_{\rm a,3} $   	                                 & \hspace{0.5cm} $-2 $     &$-(1+k)$\hspace{0.5cm}     $-(1+k)$          &$-\frac{5+\alpha_s+k(1+\alpha_s)}{\alpha_s+5-k}$\hspace{0.5cm} $-\frac{8+\alpha_s(1+k)}{\alpha_s+8-2k} $ &\hspace{0.5cm} $-\frac{\alpha_s(1+k)+2(3-k)}{\alpha_s+6-2k}$\\
$ \nu_{\rm a,3} < \nu < \nu_{\rm c} $   	                & \hspace{0.5cm} $-\frac13$   &$\frac{6k-11}{3}$\hspace{0.5cm} $\frac{6k-11}{3}$       &  $-\frac{2(5-4k)+\alpha_s(11-6k)}{3(\alpha_s+5-k)} $ \hspace{0.5cm} $\frac{6k(\alpha_s+1)-4-11\alpha_s}{3(\alpha_s+8-2k)} $	            &\hspace{0.5cm} $\frac{18-11\alpha_s+6k(\alpha_s-1)}{3(\alpha_s+6-2k)} $\\	
$\nu_{\rm c} < \nu < \nu_{\rm m} $   	                & \hspace{0.5cm} $\frac{1}{2} $  &$\frac{3k-8}{4}$\hspace{0.5cm} $\frac{3k-8}{4}$ & $-\frac{2(5-2k)+\alpha_s(8-3k)}{4(\alpha_s+5-k)} $\hspace{0.5cm} $\frac{8(1-\alpha_s)+k(3\alpha_s-2)}{4(\alpha_s+8-2k)} $	                    &\hspace{0.5cm} $\frac{8(3-k)+\alpha_s(3k-8)}{4(\alpha_s+6-2k)} $\\ 	
$\nu_{\rm m} < \nu $   	                                 & \hspace{0.5cm} $\frac{p}{2} $  &$\frac{k(2+p)-8}{4}$\hspace{0.5cm} $\frac{k(2+p)-8}{4}$  & $\frac{30p+kp(\alpha_s-8)-8(\alpha_s+5)+2k(\alpha_s+6)}{4(\alpha_s+5-k)}$  \hspace{0.5cm} $\frac{8(3p-2-\alpha_s)+kp(\alpha_s-6)+2k(2+\alpha_s)}{4(\alpha_s+8-2k)} $	                    &\hspace{0.5cm} $\frac{p[24+k(\alpha_s-8)]+2\alpha_s(k-4)}{4(\alpha_s+6-2k)}$\\ \hline
$\nu_{\rm a,1} < \nu_{\rm m} < \nu_{\rm c} $ \\\hline

$\nu < \nu_{\rm a,1} $   	                                 & \hspace{0.5cm} $-2 $    &$-2$\hspace{0.5cm}      $-2$          &$-\frac{2(\alpha_s+k-1)}{\alpha_s+5-k}$\hspace{0.5cm} $-\frac{2(\alpha_s+2)}{\alpha_s+8-2k} $	                    &\hspace{0.5cm} $-\frac{2\alpha_s}{\alpha_s+6-2k}$\\
$ \nu_{\rm a,1} < \nu < \nu_{\rm m} $   	        & \hspace{0.5cm} $-\frac{1}{3}$    &$\frac{4k-9}{3}$\hspace{0.5cm} $\frac{4k-9}{3}$     &$-\frac{3(8+3\alpha_s)-2k(2\alpha_s+5)}{3(\alpha_s+5-k)} $\hspace{0.5cm} $-\frac{6(2-k)+\alpha_s(9-4k)}{3(\alpha_s+8-2k)} $	                    &\hspace{0.5cm} $\frac{3(2-3\alpha_s)+2k(2\alpha_s-1)}{3(\alpha_s+6-2k)} $\\	
$ \nu_{\rm m} < \nu < \nu_{\rm c} $   	                & \hspace{0.5cm} $\frac{p-1}{2} $ &$\frac{k(5+p)-12}{4}$\hspace{0.5cm} $\frac{k(5+p)-12}{4}$   &$\frac{6(5p-2\alpha_s-7)+k[16+p(\alpha_s-8)+5\alpha_s]}{4(\alpha_s+5-k)}$\hspace{0.5cm} $\frac{12(2p-2-\alpha_s)+kp(\alpha_s-6)+5k(\alpha_s+2)}{4(\alpha_s+8-2k)} $	                    &\hspace{0.5cm} $\frac{p[24+k(\alpha_s-8)]+\alpha_s(5k-12)}{4(\alpha_s+6-2k)}$\\ 	
$\nu_{\rm c} < \nu $   	                                 & \hspace{0.5cm} $\frac{p}{2} $   &$\frac{k(2+p)-8}{4}$\hspace{0.5cm} $\frac{k(2+p)-8}{4}$    &$\frac{30p+kp(\alpha_s-8)-8(\alpha_s+5)+2k(\alpha_s+6)}{4(\alpha_s+5-k)}$\hspace{0.5cm} $\frac{8(3p-2-\alpha_s)+kp(\alpha_s-6)+2k(2+\alpha_s)}{4(\alpha_s+8-2k)} $	                    &\hspace{0.5cm} $\frac{p[24+k(\alpha_s-8)]+2\alpha_s(k-4)}{4(\alpha_s+6-2k)}$\\ \hline
$\nu_{\rm m} < \nu_{\rm a,2} < \nu_{\rm c} $ \\\hline 	

$\nu < \nu_{\rm m} $   	                                 & \hspace{0.5cm} $-2 $     &$-2$\hspace{0.5cm} $-2$              &$-\frac{2(\alpha_s+k-1)}{\alpha_s+5-k}$\hspace{0.5cm} $-\frac{2(\alpha_s+2)}{\alpha_s+8-2k} $	                    &\hspace{0.5cm} $-\frac{2\alpha_s}{\alpha_s+6-2k}$\\
$ \nu_{\rm m} < \nu < \nu_{\rm a,2} $   	        & \hspace{0.5cm} $-\frac52$     &$-\frac{8+k}{4}$\hspace{0.5cm} $-\frac{8+k}{4}$     &$-\frac{22+\alpha_s(8+k)}{4(\alpha_s+5-k)} $\hspace{0.5cm} $-\frac{8(\alpha_s+5)+k(\alpha_s-6)}{4(\alpha_s+8-2k)} $	                    &\hspace{0.5cm} $\frac{k(8-\alpha_s)-8(\alpha_s+3)}{4(\alpha_s+6-2k)} $\\	
$ \nu_{\rm a,2} < \nu < \nu_{\rm c} $   	                & \hspace{0.5cm} $\frac{p-1}{2} $ &$\frac{k(5+p)-12}{4}$\hspace{0.5cm} $\frac{k(5+p)-12}{4}$  &$\frac{6(5p-2\alpha_s-7)+k[16+p(\alpha_s-8)+5\alpha_s]}{4(\alpha_s+5-k)}$\hspace{0.5cm} $\frac{12(2p-2-\alpha_s)+kp(\alpha_s-6)+5k(\alpha_s+2)}{4(\alpha_s+8-2k)} $	                    &\hspace{0.5cm} $\frac{p[24+k(\alpha_s-8)]+\alpha_s(5k-12)}{4(\alpha_s+6-2k)}$\\ 	
$\nu_{\rm c} < \nu $   	                                 & \hspace{0.5cm} $\frac{p}{2} $  &$\frac{k(2+p)-8}{4}$\hspace{0.5cm}  $\frac{k(2+p)-8}{4}$  &$\frac{30p+kp(\alpha_s-8)-8(\alpha_s+5)+2k(\alpha_s+6)}{4(\alpha_s+5-k)}$\hspace{0.5cm} $\frac{8(3p-2-\alpha_s)+kp(\alpha_s-6)+2k(2+\alpha_s)}{4(\alpha_s+8-2k)} $	                    &\hspace{0.5cm} $\frac{p[24+k(\alpha_s-8)]+2\alpha_s(k-4)}{4(\alpha_s+6-2k)}$\\ \hline

%
%

\end{tabular}}
\end{table}

\clearpage
\newpage

\begin{table}
\centering \renewcommand{\arraystretch}{1.85}\addtolength{\tabcolsep}{1.5pt}
\caption{Closure relations of synchrotron afterglow radiation from a quasi-spherical material in a stratified environment}
\label{Table3}
\rotatebox{90}{\begin{tabular}{c c c  c c}
 \hline \hline
&\hspace{0.5cm}      &\hspace{0.5cm}   Coasting Phase  &Deceleration Phase &\hspace{0.5cm}   Lateral Expansion\\ 
&\hspace{0.5cm}      &Sub-relativ \hspace{0.5cm}   Relativistic     &Sub-relativ \hspace{0.5cm}   Relativistic & \hspace{0.5cm}   Relativistic\\ 
& \hspace{0.5cm} $\beta$ & \hspace{0.5cm} $\alpha(\beta)$ & \hspace{0.5cm} $\alpha(\beta)$ & \hspace{0.5cm} $\alpha(\beta)$ \\ \hline

$\nu_{\rm a,3} < \nu_{\rm c} < \nu_{\rm m} $ \\ \hline
$\nu < \nu_{\rm a,3}$   	                                 & \hspace{0.5cm} $-2$     &$\frac{(1+k)\beta}{2}$\hspace{0.5cm}     $\frac{(1+k)\beta}{2}$                 &$\frac{[5+\alpha_s+k(1+\alpha_s)]\beta}{2(\alpha_s+5-k)}$\hspace{0.5cm} $\frac{[8+\alpha_s(1+k)]\beta}{2(\alpha_s+8-2k)}$ &\hspace{0.5cm} $\frac{[\alpha_s(1+k)+2(3-k)]\beta}{2(\alpha_s+6-2k)}$\\
$ \nu_{\rm a,3} < \nu < \nu_{\rm c}$   	                & \hspace{0.5cm} $-\frac13$   &$(11-6k)\beta$\hspace{0.5cm}     $(11-6k)\beta$          &$\frac{[2(5-4k)+\alpha_s(11-6k)]\beta}{\alpha_s+5-k}$\hspace{0.5cm} $-\frac{[6k(\alpha_s+1)-4-11\alpha_s]\beta}{\alpha_s+8-2k}$	            &\hspace{0.5cm} $-\frac{[18-11\alpha_s+6k(\alpha_s-1)]\beta}{\alpha_s+6-2k}$\\	
 $\nu_{\rm c} < \nu < \nu_{\rm m}$   	                & \hspace{0.5cm}  $\frac{1}{2}$  &$\frac{(3k-8)\beta}{2}$\hspace{0.5cm}     $\frac{(3k-8)\beta}{2}$    &$-\frac{[2(5-2k)+\alpha_s(8-3k)]\beta}{2(\alpha_s+5-k)}$\hspace{0.5cm}  $\frac{[8(1-\alpha_s)+k(3\alpha_s-2)]\beta}{2(\alpha_s+8-2k)}$	                    &\hspace{0.5cm}  $\frac{[8(3-k)+\alpha_s(3k-8)]\beta}{2(\alpha_s+6-2k)}$\\ 	
 $\nu_{\rm m} < \nu$   	                                 & \hspace{0.5cm}  $\frac{p}{2}$  &$\frac{k(\beta+1)-4}{2}$\hspace{0.5cm}     $\frac{k(\beta+1)-4}{2}$     &$\frac{\beta[30+k(\alpha_s-8)]-4(\alpha_s+5)+k(\alpha_s+6)}{2(\alpha_s+5-k)}$\hspace{0.5cm}  $\frac{8(6\beta-2-\alpha_s)+2k\beta(\alpha_s-6)+2k(2+\alpha_s)}{4(\alpha_s+8-2k)}$	                    &\hspace{0.5cm}  $\frac{\beta[24+k(\alpha_s-8)]+\alpha_s(k-4)}{2(\alpha_s+6-2k)}$\\ \hline
$\nu_{\rm a,1} < \nu_{\rm m} < \nu_{\rm c}$ \\\hline

 $\nu < \nu_{\rm a,1}$   	                                 & \hspace{0.5cm}  $-2$    &$\beta$\hspace{0.5cm}     $\beta$                  & $\frac{(\alpha_s+k-1)\beta}{\alpha_s+5-k}$\hspace{0.5cm}  $\frac{(\alpha_s+2)\beta}{\alpha_s+8-2k}$	                    &\hspace{0.5cm}  $\frac{\alpha_s\beta}{\alpha_s+6-2k}$\\
 $ \nu_{\rm a,1} < \nu < \nu_{\rm m}$   	        & \hspace{0.5cm}  $-\frac{1}{3}$   &$(9-4k)\beta$\hspace{0.5cm}    $(9-4k)\beta$          &$\frac{[3(8+3\alpha_s)-2k(2\alpha_s+5)]\beta}{\alpha_s+5-k}$\hspace{0.5cm}  $\frac{[6(2-k)+\alpha_s(9-4k)]\beta}{\alpha_s+8-2k}$	                    &\hspace{0.5cm}  $-\frac{[3(2-3\alpha_s)+2k(2\alpha_s-1)]\beta}{\alpha_s+6-2k}$\\	
 $ \nu_{\rm m} < \nu < \nu_{\rm c}$   	                & \hspace{0.5cm}  $\frac{p-1}{2}$ &$\frac{k(\beta+3)-6}{2}$\hspace{0.5cm}    $\frac{k(\beta+3)-6}{2}$   &$\frac{6(5\beta-\alpha_s-1)+k[4+3\alpha_s+\beta(\alpha_s-8)]}{2(\alpha_s+5-k)}$\hspace{0.5cm}  $\frac{\beta(24+\alpha_s k-6k)+3\alpha_s(k-2)+2k}{2(\alpha_s+8-2k)}$	                    &\hspace{0.5cm}  $\frac{\beta[24+k(\alpha_s-8)]+3\alpha_s(k-2)+4(3-k)}{2(\alpha_s+6-2k)}$\\ 	
 $\nu_{\rm c} < \nu$   	                                 & \hspace{0.5cm}  $\frac{p}{2}$  &$\frac{k(\beta+1)-4}{2}$\hspace{0.5cm}     $\frac{k(\beta+1)-4}{2}$     &$\frac{\beta[30+k(\alpha_s-8)]-4(\alpha_s+5)+k(\alpha_s+6)}{2(\alpha_s+5-k)}$\hspace{0.5cm}  $\frac{8(6\beta-2-\alpha_s)+2k\beta(\alpha_s-6)+2k(2+\alpha_s)}{4(\alpha_s+8-2k)}$	                    &\hspace{0.5cm}  $\frac{\beta[24+k(\alpha_s-8)]+\alpha_s(k-4)}{2(\alpha_s+6-2k)}$\\ \hline
$\nu_{\rm m} < \nu_{\rm a,2} < \nu_{\rm c}$ \\\hline 	

 $\nu < \nu_{\rm m}$   	                                 & \hspace{0.5cm}  $-2$     &$\beta$\hspace{0.5cm}     $\beta$                 &$\frac{(\alpha_s+k-1)\beta}{\alpha_s+5-k}$\hspace{0.5cm}  $\frac{(\alpha_s+2)\beta}{\alpha_s+8-2k}$	                    &\hspace{0.5cm}  $\frac{\alpha_s\beta}{\alpha_s+6-2k}$\\
 $ \nu_{\rm m} < \nu < \nu_{\rm a,2}$   	        & \hspace{0.5cm}  $-\frac52$    &$\frac{(8+k)\beta}{10}$\hspace{0.5cm}     $\frac{(8+k)\beta}{10}$        &$\frac{[22+\alpha_s(8+k)]\beta}{10(\alpha_s+5-k)}$\hspace{0.5cm}  $\frac{[8(\alpha_s+5)+k(\alpha_s-6)]\beta}{10(\alpha_s+8-2k)}$	                    &\hspace{0.5cm}  $-\frac{[k(8-\alpha_s)-8(\alpha_s+3)]\beta}{10(\alpha_s+6-2k)}$\\	
 $ \nu_{\rm a,2} < \nu < \nu_{\rm c}$   	                & \hspace{0.5cm}  $\frac{p-1}{2}$ &$\frac{k(\beta+3)-6}{2}$\hspace{0.5cm}     $\frac{k(\beta+3)-6}{2}$   &$\frac{6(5\beta-\alpha_s-1)+k[4+3\alpha_s+\beta(\alpha_s-8)]}{2(\alpha_s+5-k)}$\hspace{0.5cm}  $\frac{\beta(24+\alpha_s k-6k)+3\alpha_s(k-2)+2k}{2(\alpha_s+8-2k)}$	                    &\hspace{0.5cm}  $\frac{\beta[24+k(\alpha_s-8)]+3\alpha_s(k-2)+4(3-k)}{2(\alpha_s+6-2k)}$\\ 	
 $\nu_{\rm c} < \nu$   	                                 & \hspace{0.5cm}  $\frac{p}{2}$  &$\frac{k(\beta+1)-4}{2}$\hspace{0.5cm}     $\frac{k(\beta+1)-4}{2}$     &$\frac{\beta[30+k(\alpha_s-8)]-4(\alpha_s+5)+k(\alpha_s+6)}{2(\alpha_s+5-k)}$\hspace{0.5cm}  $\frac{8(6\beta-2-\alpha_s)+2k\beta(\alpha_s-6)+2k(2+\alpha_s)}{4(\alpha_s+8-2k)}$	                    &\hspace{0.5cm}  $\frac{\beta[24+k(\alpha_s-8)]+\alpha_s(k-4)}{2(\alpha_s+6-2k)}$\\ \hline 	
\end{tabular}}
\end{table}

\begin{table}
\centering \renewcommand{\arraystretch}{1.85}\addtolength{\tabcolsep}{1.5pt}
\caption{Evolution of the density parameter $F_\nu\propto A_{\rm k}^{\alpha_{\rm k}}$ in each cooling condition of the synchrotron afterglow model} \label{TableDensityParameter}
\begin{tabular}{c c c  c c}
 \hline \hline
&\hspace{0.5cm}      &\hspace{0.5cm}   Coasting Phase  &Deceleration Phase &\hspace{0.5cm}   Lateral Expansion\\ 
&\hspace{0.5cm}      &Sub-relativistic \hspace{0.5cm}   Relativistic     &Sub-relativistic \hspace{0.5cm}   Relativistic & \hspace{0.5cm}   Relativistic\\ 
& \hspace{0.5cm} $\beta$ & \hspace{0.5cm} $\alpha(\beta)$ & \hspace{0.5cm} $\alpha(\beta)$ & \hspace{0.5cm} $\alpha(\beta)$ \\ \hline

$ $                     & \hspace{0.5cm}  $\beta $            &   \hspace{0.5cm}  $\alpha_{\rm k} $  &  \hspace{0.5cm}  $\alpha_{\rm k} $\\  \hline \hline
$\nu_{\rm a,3} < \nu_{\rm c} < \nu_{\rm m} $ \\ \hline

$\nu < \nu_{\rm a,3} $   	                                 & \hspace{0.5cm} $-2 $  &$-1$\hspace{0.5cm}  $-1$                   &$-\frac{5+\alpha_s}{\alpha_s+5-k}$\hspace{0.5cm} $-\frac{\alpha_s+8}{\alpha_s+8-2k} $ &\hspace{0.5cm} $-\frac{6+\alpha_s}{\alpha_s+6-2k}$\\
$ \nu_{\rm a,3} < \nu < \nu_{\rm c} $   	                & \hspace{0.5cm} $-\frac13$   &$2$\hspace{0.5cm} $2$          &$\frac{5+2\alpha_s}{\alpha_s+5-k} $\hspace{0.5cm} $\frac{2(10+3\alpha_s)}{3(\alpha_s+8-2k)} $	            &\hspace{0.5cm} $\frac{2(4+3\alpha_s)}{3(\alpha_s+6-2k)} $\\	
$\nu_{\rm c} < \nu < \nu_{\rm m} $   	                & \hspace{0.5cm} $\frac{1}{2} $  &$\frac34$\hspace{0.5cm}  $\frac34$   &$\frac{5+3\alpha_s}{4(\alpha_s+5-k)} $\hspace{0.5cm} $\frac{3\alpha_s}{4(\alpha_s+8-2k)} $	                    &\hspace{0.5cm} $\frac{3(\alpha_s-2)}{4(\alpha_s+6-2k)} $\\ 	
$\nu_{\rm m} < \nu $   	                                 & \hspace{0.5cm} $\frac{p}{2} $ &$\frac{p+2}{4}$\hspace{0.5cm}  $\frac{p+2}{4}$     &$\frac{2(5+\alpha_s)+p(\alpha_s-5)}{4(\alpha_s+5-k)} $\hspace{0.5cm} $\frac{(p+2)\alpha_s}{4(\alpha_s+8-2k)} $	                    &\hspace{0.5cm} $\frac{(p+2)(\alpha_s-2)}{4(\alpha_s+6-2k)} $\\ \hline
$\nu_{\rm a,1} < \nu_{\rm m} < \nu_{\rm c} $ \\\hline

$\nu < \nu_{\rm a,1} $   	                                 & \hspace{0.5cm} $-2 $    &$0$\hspace{0.5cm}    $0$               &$-\frac{4}{\alpha_s+5-k}$\hspace{0.5cm} $-\frac{4}{\alpha_s+8-2k} $	                    &\hspace{0.5cm} $-\frac{4}{\alpha_s+6-2k}$\\
$ \nu_{\rm a,1} < \nu < \nu_{\rm m} $   	        & \hspace{0.5cm} $-\frac{1}{3}$    &$\frac43$\hspace{0.5cm}   $\frac43$       &$\frac{13+4\alpha_s}{3(\alpha_s+5-k)} $\hspace{0.5cm} $\frac{4(\alpha_s+3)}{3(\alpha_s+8-2k)} $	                    &\hspace{0.5cm} $\frac{4(\alpha_s+1)}{3(\alpha_s+6-2k)} $\\	
$ \nu_{\rm m} < \nu < \nu_{\rm c} $   	                & \hspace{0.5cm} $\frac{p-1}{2} $ &$\frac{p+5}{4}$\hspace{0.5cm} $\frac{p+5}{4}$   &$\frac{19+5\alpha_s+p(\alpha_s-5)}{4(\alpha_s+5-k)} $\hspace{0.5cm} $\frac{16+(p+5)\alpha_s}{4(\alpha_s+8-2k)} $	                    &\hspace{0.5cm} $\frac{6+5\alpha_s+p(\alpha_s-2)}{4(\alpha_s+6-2k)} $\\ 	
$\nu_{\rm c} < \nu $   	                                 & \hspace{0.5cm} $\frac{p}{2} $  &$\frac{p+2}{4}$\hspace{0.5cm}  $\frac{p+2}{4}$    &$\frac{2(5+\alpha_s)+p(\alpha_s-5)}{4(\alpha_s+5-k)} $\hspace{0.5cm} $\frac{(p+2)\alpha_s}{4(\alpha_s+8-2k)} $	                    &\hspace{0.5cm} $\frac{(p+2)(\alpha_s-2)}{4(\alpha_s+6-2k)} $\\ \hline
$\nu_{\rm m} < \nu_{\rm a,2} < \nu_{\rm c} $ \\\hline 	

$\nu < \nu_{\rm m} $   	                                 & \hspace{0.5cm} $-2 $       &$0$\hspace{0.5cm}  $0$              &$-\frac{4}{\alpha_s+5-k}$\hspace{0.5cm} $-\frac{4}{\alpha_s+8-2k} $	                    &\hspace{0.5cm} $-\frac{4}{\alpha_s+6-2k}$\\
$ \nu_{\rm m} < \nu < \nu_{\rm a,2} $   	        & \hspace{0.5cm} $-\frac52$     &$-\frac14$\hspace{0.5cm}   $-\frac14$      &$-\frac{11+\alpha_s}{4(\alpha_s+5-k)} $\hspace{0.5cm} $-\frac{16+\alpha_s}{4(\alpha_s+8-2k)} $	                    &\hspace{0.5cm} $-\frac{14+\alpha_s}{4(\alpha_s+6-2k)} $\\	
$ \nu_{\rm a,2} < \nu < \nu_{\rm c} $   	                & \hspace{0.5cm} $\frac{p-1}{2} $ &$\frac{p+5}{4}$\hspace{0.5cm} $\frac{p+5}{4}$   &$\frac{19+5\alpha_s+p(\alpha_s-5)}{4(\alpha_s+5-k)} $\hspace{0.5cm} $\frac{16+(p+5)\alpha_s}{4(\alpha_s+8-2k)} $	                    &\hspace{0.5cm} $\frac{6+5\alpha_s+p(\alpha_s-2)}{4(\alpha_s+6-2k)} $\\ 	
$\nu_{\rm c} < \nu $   	                                 & \hspace{0.5cm} $\frac{p}{2} $  &$\frac{p+2}{4}$\hspace{0.5cm}  $\frac{p+2}{4}$    &$\frac{2(5+\alpha_s)+p(\alpha_s-5)}{4(\alpha_s+5-k)} $\hspace{0.5cm} $\frac{(p+2)\alpha_s}{4(\alpha_s+8-2k)} $	                    &\hspace{0.5cm} $\frac{(p+2)(\alpha_s-2)}{4(\alpha_s+6-2k)} $\\ \hline

%
%
\hline
\end{tabular}
\end{table}

\clearpage

\begin{table*}
 \centering \renewcommand{\arraystretch}{2.5}\addtolength{\tabcolsep}{2pt}
 \caption{The best-fit values found with MCMC simulations after describing the multiwavelength afterglow observations with a synchrotron model evolving in a constant medium.}
 \label{best_fit_par}
 \begin{tabular}{l c c c c c c c c}
 \hline
  GRB Name & $\mathrm{log_{10}(E/erg)}$ & $\mathrm{log_{10}(n/cm^{-3})}$ & $\Gamma_{0}$ & $\mathrm{log_{10}(\varepsilon_{e})}$ & $\mathrm{log_{10}(\varepsilon_{B})}$  & $\mathrm{p}$ & $\mathrm{\alpha}$ & $\mathrm{\theta_{c}}$ \\ \hline

 980425  & $51.993_{-0.010}^{+0.009}$ & $-0.119_{-0.008}^{+0.009}$ & $3.758_{-0.011}^{+0.012}$ & $-1.023_{-0.010}^{+0.011}$ & $-1.478_{-0.011}^{+0.010}$ & $2.803_{-0.010}^{+0.009}$ & $2.000_{-0.009}^{+0.011}$ & $50.305_{-0.011}^{+0.011}$ \\
 031203  & $52.000_{-0.010}^{+0.010}$ & $0.254_{-0.010}^{+0.009}$ & $2.870_{-0.009}^{+0.012}$ & $-0.110_{-0.010}^{+0.010}$ & $-1.001_{-0.010}^{+0.010}$ & $2.059_{-0.011}^{+0.010}$ & $3.001_{-0.011}^{+0.008}$ & $32.019_{-0.009}^{+0.011}$ \\
 060218  & $49.751_{-0.010}^{+0.010}$ & $0.473_{-0.010}^{+0.010}$ & $2.111_{-0.009}^{+0.011}$ & $-0.264_{-0.010}^{+0.010}$ & $-4.772_{-0.010}^{+0.010}$ & $3.030_{-0.010}^{+0.009}$ & $2.999_{-0.010}^{+0.009}$ & $46.009_{-0.010}^{+0.010}$ \\
 100316D & $51.994_{-0.010}^{+0.010}$ & $0.756_{-0.009}^{+0.010}$ & $1.703_{-0.010}^{+0.010}$ & $-2.034_{-0.011}^{+0.010}$ & $-1.663_{-0.011}^{+0.009}$ & $2.500_{-0.010}^{+0.009}$ & $2.999_{-0.010}^{+0.011}$ & $51.932_{-0.010}^{+0.009}$ \\
130603B & $51.474_{-0.011}^{+0.009}$ & $0.010_{-0.010}^{+0.010}$ & $4.799_{-0.009}^{+0.012}$ & $-0.797_{-0.010}^{+0.010}$ & $-1.981_{-0.010}^{+0.010}$ & $2.027_{-0.010}^{+0.010}$ & $2.301_{-0.011}^{+0.009}$ & $44.963_{-0.010}^{+0.010}$ \\
150101B & $51.992_{-0.010}^{+0.011}$ & $-0.001_{-0.009}^{+0.010}$ & $4.480_{-0.010}^{+0.010}$ & $-0.393_{-0.010}^{+0.011}$ & $-1.934_{-0.009}^{+0.011}$ & $2.533_{-0.009}^{+0.010}$ & $2.163_{-0.011}^{+0.010}$ & $43.981_{-0.010}^{+0.010}$ \\
 171205A & $52.000_{-0.002}^{+0.002}$ & $1.000_{-0.002}^{+0.002}$ & $4.894_{-0.002}^{+0.002}$ & $-0.097_{-0.002}^{+0.002}$ & $-2.000_{-0.002}^{+0.002}$ & $2.266_{-0.002}^{+0.002}$ & $2.000_{-0.002}^{+0.002}$ & $54.844_{-0.002}^{+0.002}$ \\
 
\hline
 \end{tabular}
\end{table*}

\begin{table*}
 \centering \renewcommand{\arraystretch}{2.5}\addtolength{\tabcolsep}{2pt}
 \caption{The best-fit values found with MCMC simulations after describing the multiwavelength afterglow observations with a synchrotron model evolving in a wind-like medium.}
 \label{best_fit_par_wind}
 \begin{tabular}{l c c c c c c c c}
 \hline
  GRB Name & $\mathrm{log_{10}(E/erg)}$ & $\mathrm{log_{10}(n/cm^{-3})}$ & $\Gamma_{0}$ & $\mathrm{log_{10}(\varepsilon_{e})}$ & $\mathrm{log_{10}(\varepsilon_{B})}$  & $\mathrm{p}$ & $\mathrm{\alpha_{\rm s}}$ & $\mathrm{\theta_{c}}$ \\ \hline

 980425  & $45.035_{-0.026}^{+0.058}$ & $-1.737_{-0.196}^{+0.416}$ & $5.506_{-3.064}^{+3.071}$ & $-0.261_{-0.053}^{+0.054}$ & $-0.417_{-0.173}^{+0.089}$ & $2.434_{-0.003}^{+0.004}$ & $2.004_{-0.003}^{+0.006}$ & $42.529_{-8.505}^{+8.482}$ \\
 031203  & $47.600_{-0.108}^{+0.110}$ & $0.827_{-0.090}^{+0.099}$ & $5.477_{-0.093}^{+0.112}$ & $-0.519_{-0.089}^{+0.090}$ & $-0.543_{-0.091}^{+0.109}$ & $2.484_{-0.091}^{+0.106}$ & $2.013_{-0.100}^{+0.089}$ & $42.473_{-0.096}^{+0.109}$ \\
 060218  & $50.035_{-0.256}^{+0.288}$ & $0.789_{-0.344}^{+0.157}$ & $5.533_{-3.057}^{+3.047}$ & $-0.518_{-0.029}^{+0.013}$ & $-0.557_{-0.094}^{+0.043}$ & $2.670_{-0.008}^{+0.011}$ & $2.022_{-0.017}^{+0.037}$ & $42.547_{-8.484}^{+8.508}$ \\
 100316D & $50.845_{-0.507}^{+0.517}$ & $0.607_{-0.615}^{+0.293}$ & $5.534_{-3.090}^{+3.029}$ & $-0.533_{-0.052}^{+0.024}$ & $-0.605_{-0.172}^{+0.078}$ & $2.667_{-0.015}^{+0.019}$ & $2.047_{-0.035}^{+0.078}$ & $42.675_{-8.608}^{+8.440}$ \\
 171205A & $51.895_{-0.165}^{+0.078}$ & $-1.668_{-2.164}^{+1.858}$ & $5.494_{-3.052}^{+3.066}$ & $-1.417_{-0.208}^{+0.241}$ & $-3.677_{-0.239}^{+0.470}$ & $2.845_{-0.008}^{+0.008}$ & $2.887_{-0.050}^{+0.052}$ & $42.527_{-8.535}^{+8.475}$ \\
\hline
 \end{tabular}
\end{table*}

\begin{figure}
{\centering
\resizebox*{0.5\textwidth}{0.33\textheight}
{\includegraphics{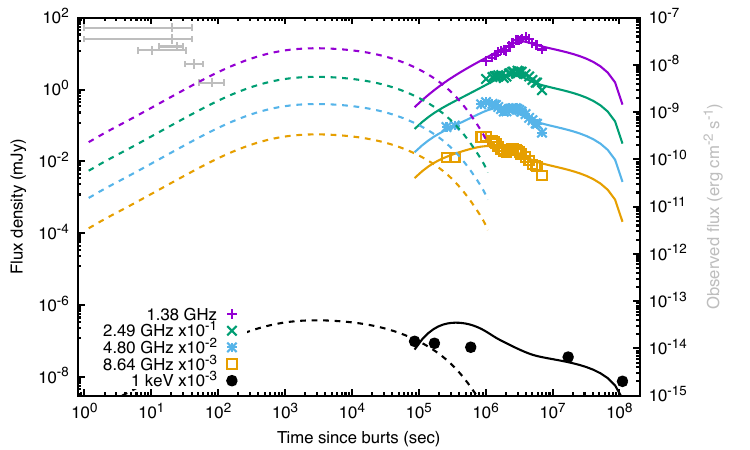}}
\resizebox*{0.5\textwidth}{0.33\textheight}
{\includegraphics{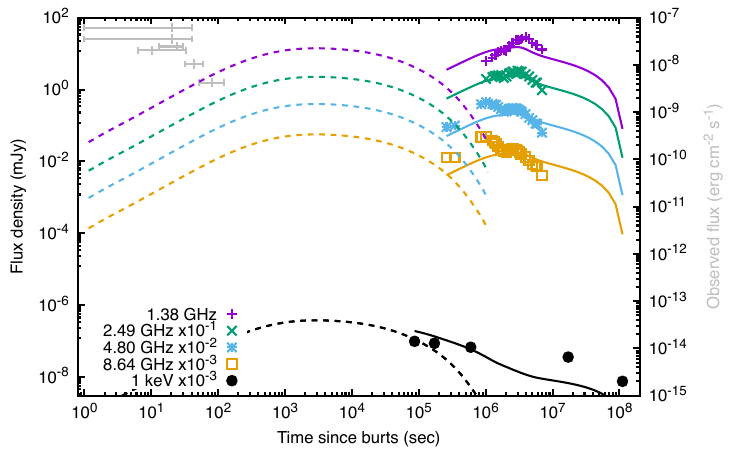}}
} 
\caption{Multiwavelength observations of GRB 980425 with the best-fit curves obtained with the synchrotron afterglow model of quasi-spherical outflow (solid line) evolving in a constant (left) and wind-like (right) medium. The {dashed} line corresponds to a narrow jet introduced to describe the early X-ray observations. Details of the narrow jet can be found in \citep{2019ApJ...884...71F}. Radio data taken from \citet{1998Natur.395..663K}, X-Ray data taken from \citet{2000ApJ...536..778P, 2004AdSpR..34.2711P, 2004ApJ...608..872K}.}
\label{lc_GRB980425}
\end{figure}

\begin{figure}
{\centering
\resizebox*{0.5\textwidth}{0.33\textheight}
{\includegraphics{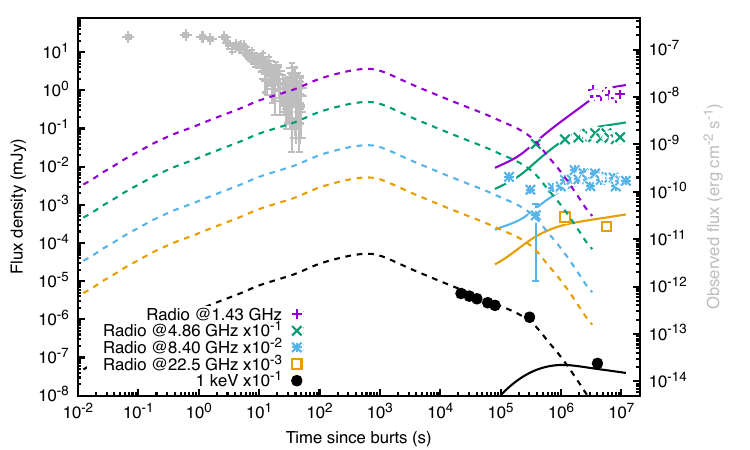}}
\resizebox*{0.5\textwidth}{0.33\textheight}
{\includegraphics{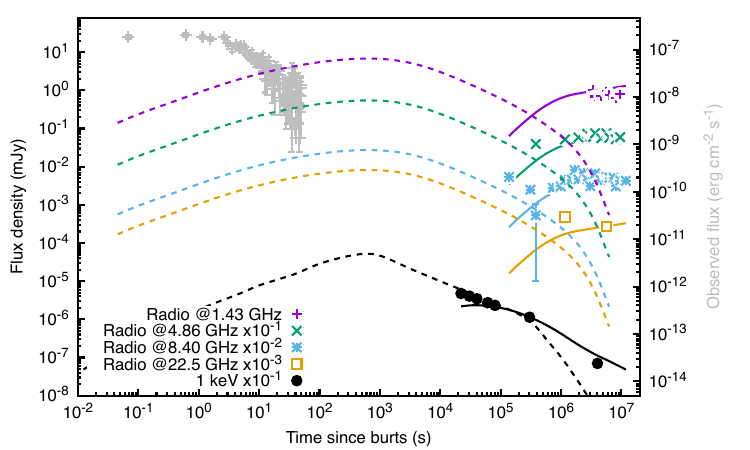}}
} 
\caption{Same as Figure \ref{lc_GRB980425}, but for GRB 031203. Radio data taken from \citet{2004Natur.430..648S}, X-Ray data taken from \citet{2004ApJ...605L.101W}.}
\label{lc_GRB031203}
\end{figure}

\begin{figure}
{\centering
\resizebox*{0.5\textwidth}{0.33\textheight}
{\includegraphics{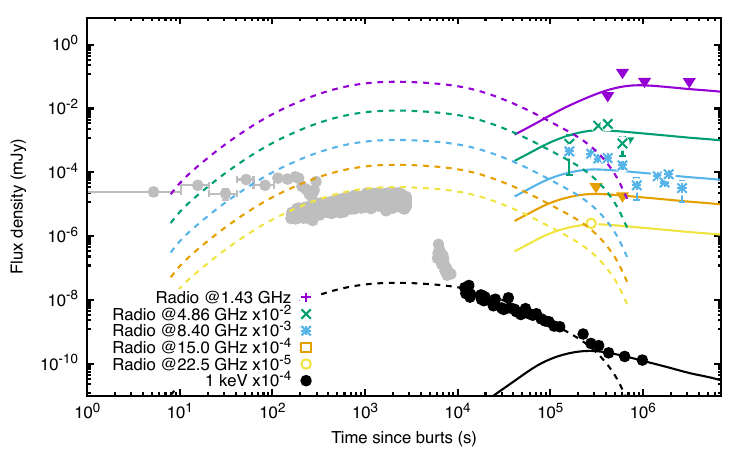}}
\resizebox*{0.5\textwidth}{0.33\textheight}
{\includegraphics{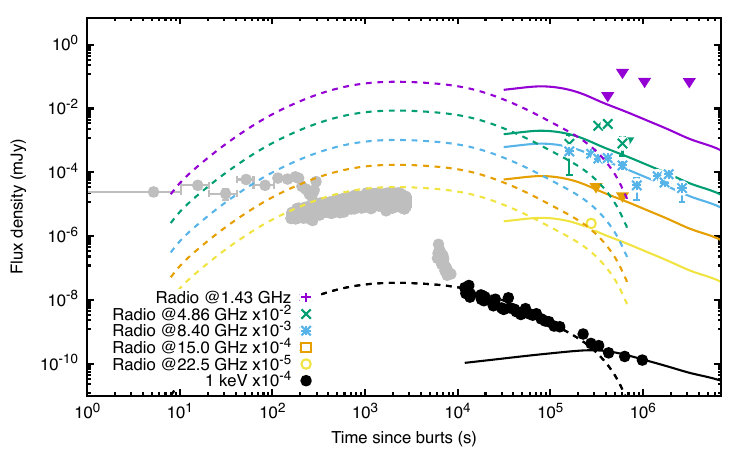}}
} 
\caption{Same as Figure \ref{lc_GRB980425}, but for GRB 060218. Radio data taken from \citet{2006Natur.442.1014S}, X-Ray data taken from \citet{2006Natur.442.1008C}.}
\label{lc_GRB060218}
\end{figure}

\begin{figure}
{\centering
\resizebox*{0.5\textwidth}{0.33\textheight}
{\includegraphics{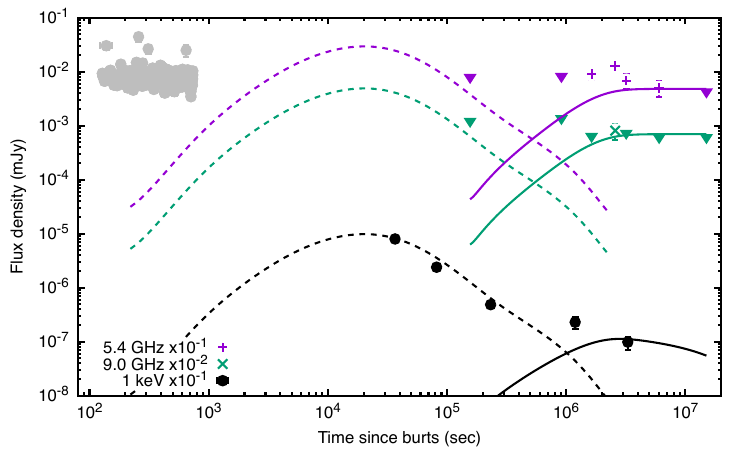}}
\resizebox*{0.5\textwidth}{0.33\textheight}
{\includegraphics{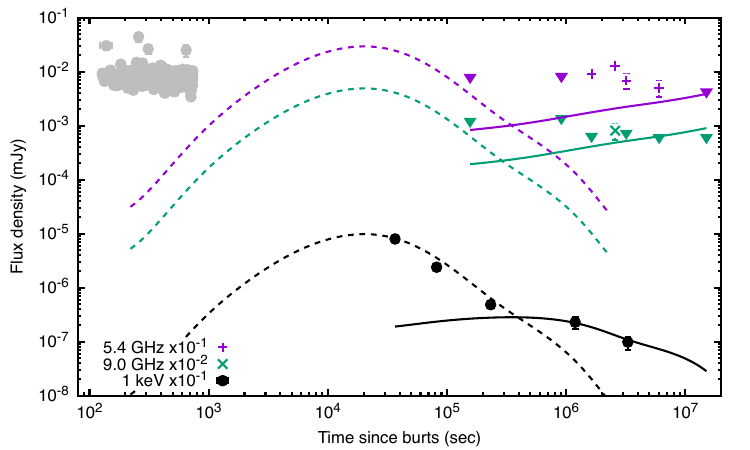}}
} 
\caption{Same as Figure \ref{lc_GRB980425}, but for GRB 100316D. Data taken from \citet{2013ApJ...778...18M}.}
\label{lc_GRB100316D}
\end{figure} 

\begin{figure}
{\centering
\resizebox*{0.5\textwidth}{0.33\textheight}
{\includegraphics{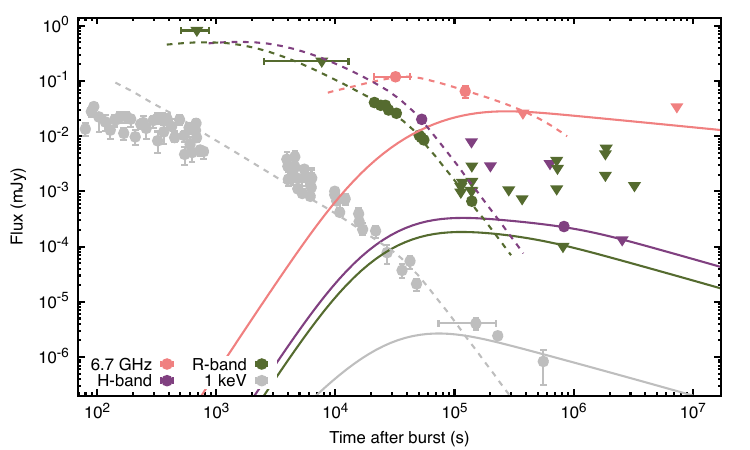}}
\resizebox*{0.5\textwidth}{0.33\textheight}
{\includegraphics{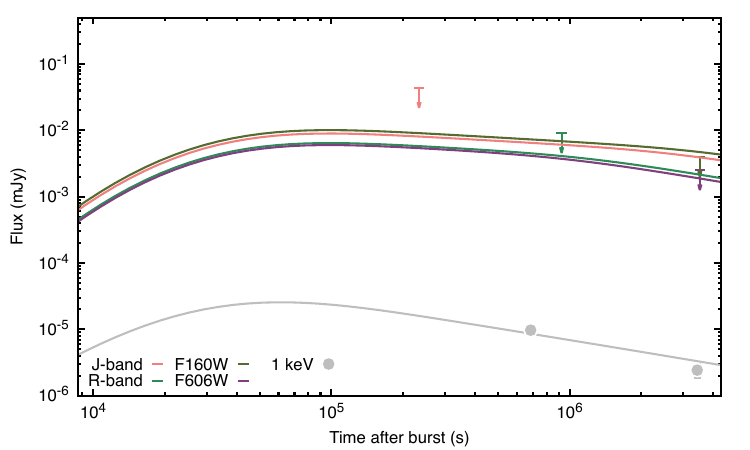}}
} 
\caption{Same as Figure \ref{lc_GRB980425}, but for GRB 130603B and GRB 150101B. For GRB 130603B  the radio data was taken from \citet{2014ApJ...780..118F, 2013ApJ...774L..23B, 2013Natur.500..547T}, optical data was taken from \citet{2013ApJ...777...94C, 2013Natur.500..547T, 2014ApJ...780..118F, 2014A&A...563A..62D} and X-Ray data from \citet{2014ApJ...780..118F}. Meanwhile, for GRB 150101B  data was taken from \citet{2016ApJ...833..151F}.}
\label{GRBs_130603B-150101B_lc_ISM}
\end{figure}

\begin{figure}
{\centering
\resizebox*{0.5\textwidth}{0.33\textheight}
{\includegraphics{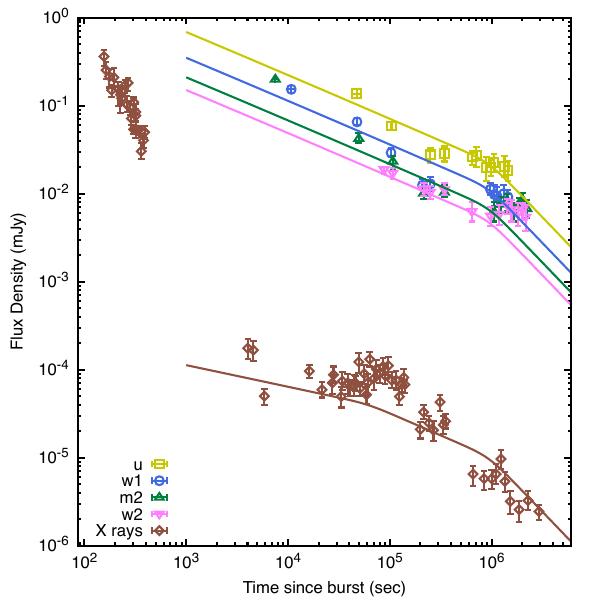}}
\resizebox*{0.5\textwidth}{0.33\textheight}
{\includegraphics{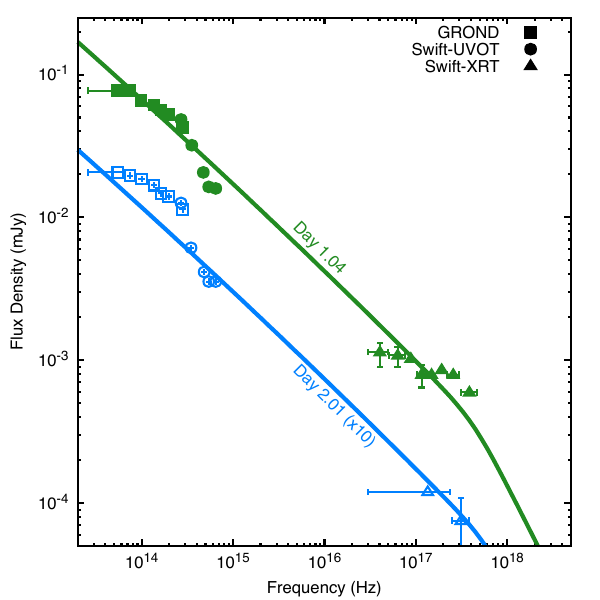}}
} 
\caption{Multiwavelength light curves (left) and SED (right) observations of GRB 171205A with the best-fit curves obtained with the synchrotron afterglow model evolving in a wind-like medium. Data taken from \citet{2019Natur.565..324I}.}
\label{GRBs_171205_lc_wind}
\end{figure}

\begin{figure}
{\centering
\resizebox*{0.5\textwidth}{0.33\textheight}
{\includegraphics{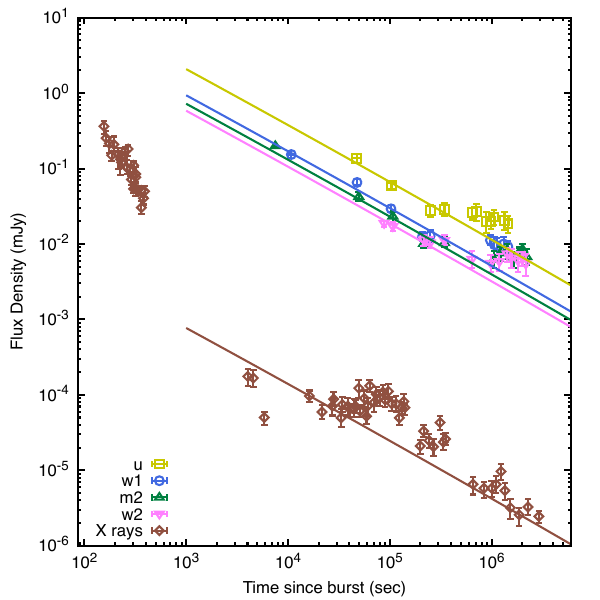}}
\resizebox*{0.5\textwidth}{0.33\textheight}
{\includegraphics{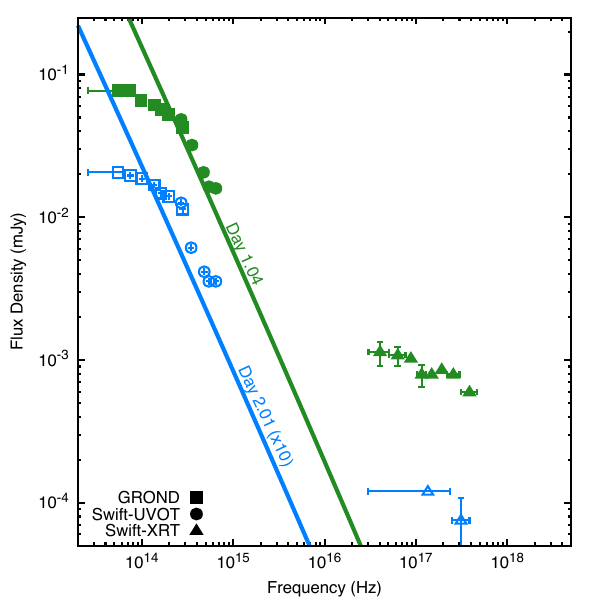}}
} 
\caption{Multiwavelength light curves (left) and SED (right) observations of GRB 171205A with the best-fit curves obtained with the synchrotron afterglow model for a constant cirbuburst like medium. Data taken from \citet{2019Natur.565..324I}.}
\label{GRBs_171205_lc_ISM}
\end{figure}


\begin{figure}
{\centering
\resizebox*{0.97\textwidth}{0.97\textheight}
{\includegraphics{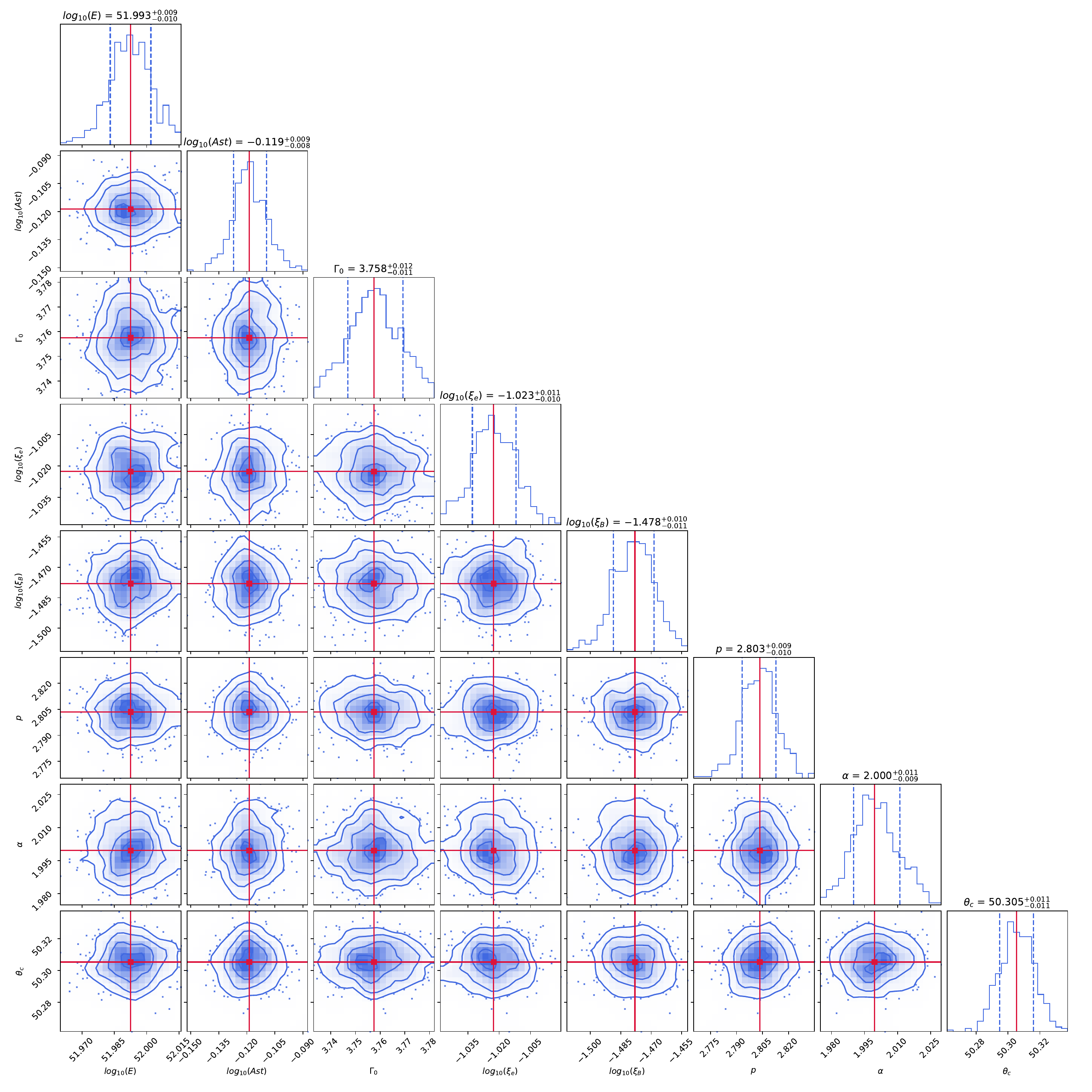}}
} 
\caption{Corner plot of the parameters obtained from describing the multiwavelength observations of GRB 980425. Red lines show the best-fit values which are reported in Table \ref{best_fit_par}. }
\label{mcmc_GRB980425}
\end{figure}

\begin{figure}
{\centering
\resizebox*{\textwidth}{0.35\textheight}
{\includegraphics{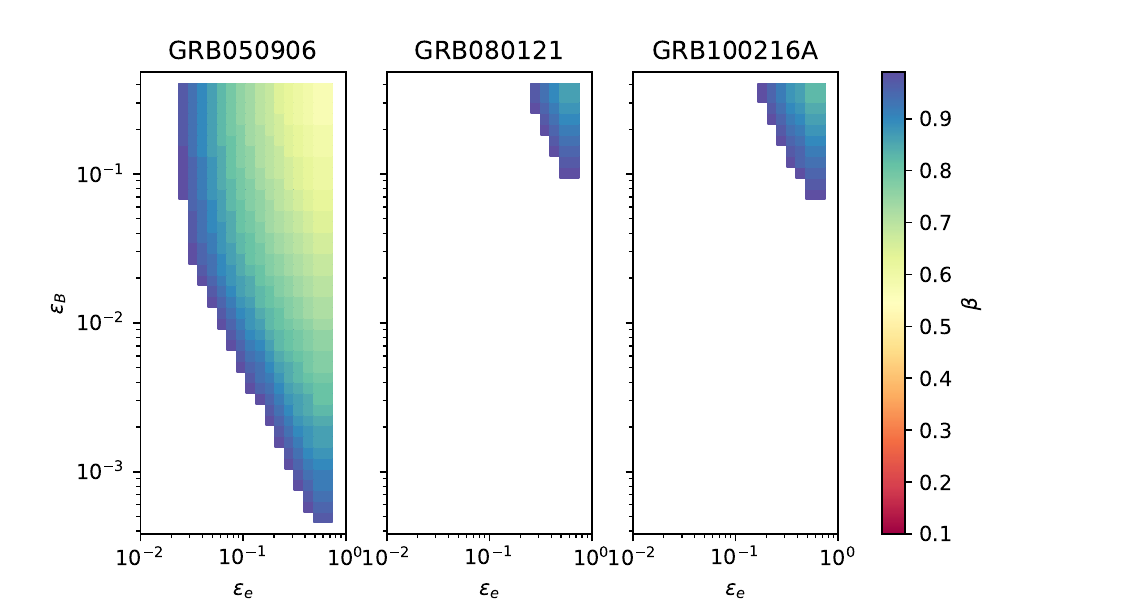}}
} 
\caption{The 3D parameter space of the rejected values of $\varepsilon_B$ as a function of $\beta$ and $\varepsilon_e$ for $n=1\,{\rm cm^{-3}}$, $E=10^{50}\,{\rm erg}$, $\alpha_s=2$ and $p=2.2$.}
\label{fig:SSGRBs}
\end{figure}

\newpage
\clearpage
\appendix
\section{synchrotron forward-shock scenario from a {cocoon} material }\label{Appendix}

\subsection{Mildly-relativistic regime}

Hereafter, we report the proportionality constants with the values of ${k=2}$ for  $n=A_{\rm st}\,3\times 10^{35}\,{\rm cm^{-1}}r^{-2}$, $p=2.4$ and $\alpha_s=2$. 

\paragraph{Coasting phase}

A constant bulk Lorentz factor characterizes the coasting phase. For the lowest energy electrons and for those with energy above which they cool effectively, the Lorentz factors are

{\small
\bary\label{eLor_syn_ism_Coast}
\gamma_{\rm m}&=& 1.65\times 10^{2}\, g(p) \varepsilon_{\rm e,-1}\zeta_{\rm e}^{-1} \, \Gamma_{0,0.5},\cr
\gamma_{\rm c}&=&  8.79\left(\frac{1+z}{1.05}\right)^{1-k} (1+Y)^{-1}\, \varepsilon_{\rm B,-3}^{-1}\,A^{-1}_{\rm st}\,\Gamma_{0,0.5}^{2k-3}\,t^{k-1}_{3}\,.
\eary
}

The characteristic and cooling synchrotron breaks and the maximum synchrotron flux become

{\small
\bary\label{En_br_syn_Coast}
\nu_{\rm m}&\simeq& 8.57\times 10^{12}\,{\rm Hz}\, \left(\frac{1+z}{1.05}\right)^{\frac{k-2}{2}}\zeta_{\rm e}^{-2}g^2(p)\varepsilon_{\rm e,-1}^2 \varepsilon_{\rm B,-3}^{\frac{1}{2}} A^{\frac12}_{\rm st} \Gamma_{0,0.5}^{4-k}\,t^{-\frac{k}{2}}_3,\cr
\nu_{\rm c}&\simeq& 3.55\times 10^{11}\,{\rm Hz} \left(\frac{1+z}{1.05}\right)^{\frac{2-3k}{2}} (1+Y)^{-2} \varepsilon_{\rm B,-3}^{-\frac{3}{2}}\,A^{-\frac32}_{\rm st} \Gamma_{0,0.5}^{3k-4}\,t^{\frac{3k-4}{2}}_3,\cr 
F_{\rm max} &\simeq& 4.80\,{\rm mJy}\left(\frac{1+z}{1.05}\right)^{\frac{3k-2}{2}} \zeta_{\rm e}\,  \varepsilon_{\rm B,-3}^{\frac{1}{2}}d_{z, 26.78}^{-2}\,A^{\frac{3}{2}}_{\rm st} \Gamma_{0,0.5}^{8-3k}\,t^{\frac{3(2-k)}{2}}_3\,.
\eary
}

The synchrotron spectral breaks in the self-absorption regime can be expressed as

{\small
\bary\label{Self_abs_syn_Coast}
\nu_{\rm a,1}&\simeq& 3.45\times 10^{4}\,{\rm Hz}\,\left(\frac{1+z}{1.05}\right)^{\frac{4(k-2)}{5}} \zeta_{\rm e}\,g(p)^{-1}\,\varepsilon_{\rm e,-1}^{-1} \varepsilon_{\rm B,-3}^{\frac{1}{5}}\,A^{\frac{4}{5}}_{\rm st}\Gamma_{0,0.5}^{\frac{8(1-k)}{5}}\,t^{\frac{3-4k}{5}}_3, \cr
\nu_{\rm a,2}&\simeq& 3.63\times 10^{8}\,{\rm Hz}\,\left(\frac{1+z}{1.05}\right)^{\frac{(k-2)(p+6)}{2(p+4)}}\zeta_{\rm e}^{-\frac{2(p-1)}{p+4}}g(p)^{\frac{2(p-1)}{p+4}}\varepsilon_{\rm B,-3}^{\frac{p+2}{2(p+4)}} \varepsilon_{\rm e,-1}^{\frac{2(p-1)}{p+4}}\, A^{\frac{p+6}{2(p+4)}}_{\rm st} \Gamma_{0,0.5}^{\frac{4(p+2)-k(p+6)}{p+4}}\, t^{\frac{4-k(p+6)}{2(p+4)}}_3,\cr 
\nu_{\rm a,3} &\simeq& 9.48\times 10^{6}\,{\rm Hz}\,\left(\frac{1+z}{1.05}\right)^{\frac{13-9k}{5}} (1+Y) \varepsilon_{\rm B,-3}^{\frac{6}{5}}\, A^{\frac{9}{5}}_{\rm st} \Gamma_{0,0.5}^{\frac{2(14-9k)}{5}}\,t^{\frac{8-9k}{5}}_3\,.
\eary
}

Table~\ref{Table2} shows the evolution of synchrotron light curves across the spectral and temporal indices, given the spectral breaks and maximum flux from eqs. \ref{En_br_syn_Coast} and \ref{Self_abs_syn_Coast}. Also presented in Table~\ref{Table3} are the closure relations for each cooling condition seen during the coasting phase.

\paragraph{Deceleration phase}

During the deceleration phase, the Lorentz factor of the lowest energy electrons and the one of those with energy above which they cool effectively are

{\small
\bary\label{eLor_syn_ism}
\gamma_{\rm m}&=& 1.41\times 10^{2}\,\left(\frac{1+z}{1.05}\right)^{-\frac{k-3}{\alpha_s+8-2k}}\, \zeta_{\rm e}^{-1}\, g(p)\, \varepsilon_{\rm e,-1} \, A^{-\frac{1}{\alpha_s+8-2k}}_{\rm st}\, {\tilde{E}_{50}}^{\frac{1}{\alpha_s+8-2k}} \,t^{\frac{k-3}{\alpha_s+8-2k}}_4,\cr
\gamma_{\rm c}&=& 7.48\times 10^{1}\,\left(\frac{1+z}{1.05}\right)^{-\frac{k+1+\alpha_s(k-1)}{\alpha_s+8-2k}} (1+Y)^{-1}\, \varepsilon_{\rm B,-3}^{-1}\,A^{-\frac{\alpha_s+5}{\alpha_s+8-2k}}_{\rm st}\,\tilde{E}_{50}^{\frac{2k-3}{\alpha_s+8-2k}}\,t^{\frac{1-\alpha_s+k(\alpha_s+1)}{\alpha_s+8-2k}}_{4}\,.
\eary
}

The characteristic and cooling synchrotron breaks and the maximum synchrotron flux become

{\small
\bary\label{En_br_syn_ism}
\nu_{\rm m}&\simeq& 6.21\times 10^{11}\,{\rm Hz}\,\left(\frac{1+z}{1.05}\right)^{\frac{8+\alpha_s(k-2)-2k}{2(\alpha_s+8-2k)}}\zeta_{\rm e}^{-2}g^2(p)\varepsilon_{\rm e,-1}^2 \varepsilon_{\rm B,-3}^{\frac{1}{2}} A^{\frac{\alpha_s}{16+2\alpha_s-4k}}_{\rm st} \tilde{E}_{50}^{\frac{4-k}{\alpha_s+8-2k}}t^{-\frac{24+(\alpha_s-6)k}{2(\alpha_s+8-2k)}}_4,\cr
\nu_{\rm c}&\simeq& 4.46\times 10^{12}\,{\rm Hz}\, \left(\frac{1+z}{1.05}\right)^{-\frac{8-2\alpha_s+2k+3\alpha_s k}{2(\alpha_s+8-2k)}}  \varepsilon_{\rm B,-3}^{-\frac{3}{2}} (1+Y)^{-2} A^{-\frac{16+3\alpha_s}{2(\alpha_s+8-2k)}}_{\rm st}\tilde{E}_{50}^{\frac{3k-4}{\alpha_s+8-2k}}t^{\frac{(\alpha_s+2)(3k-4)}{2(\alpha_s+8-2k)}}_4,\cr 
F_{\rm max} &\simeq& 3.48\,{\rm mJy}\,\left(\frac{1+z}{1.05}\right)^{\frac{32+3k(\alpha_s-2)-2\alpha_s}{2(\alpha_s+8-2k)}} \zeta_{\rm e}\, \varepsilon_{\rm B,-3}^{\frac{1}{2}}d_{z,26.78}^{-2} A^{\frac{3\alpha_s+8}{2(\alpha_s+8-2k)}}_{\rm st} \tilde{E}_{50}^{\frac{8-3k}{\alpha_s+8-2k}}t^{-\frac{3\alpha_s(k-2)+2k}{2(\alpha_s+8-2k)}}_4\,.
\eary
}

The synchrotron spectral breaks in the self-absorption regime can be expressed as

{\small
\bary\label{Self_abs_syn_ism}
\nu_{\rm a,1}&\simeq& 4.46\times 10^{3}\,{\rm Hz}\, \left(\frac{1+z}{1.05}\right)^{\frac{4k(\alpha_s+4)-8(\alpha_s+5)}{5(\alpha_s+8-2k)}} \zeta_{\rm e}\,g(p)^{-1}\varepsilon_{\rm e,-1}^{-1} \varepsilon_{\rm B,-3}^{\frac{1}{5}} A^{\frac{4(\alpha_s+6)}{5(\alpha_s+8-2k)}}_{\rm k} \tilde{E}_{50}^{\frac{8(1-k)}{5(\alpha_s+8-2k)}}t^{\frac{3\alpha_s-2k(2\alpha_s+3)}{5(\alpha_s+8-2k)}}_4,\cr
\nu_{\rm a,2}&\simeq& 3.56\times 10^{7}\,{\rm Hz}\, \left(\frac{1+z}{1.05}\right)^{\frac{k[20+p(\alpha_s-2)+6\alpha_s]-2[p(\alpha_s-4)+6(\alpha_s+4)]}{2(p+4)(\alpha_s+8-2k)}}\zeta_{\rm e}^{-\frac{2(p-1)}{p+4}}g(p)^{\frac{2(p-1)}{p+4}}\varepsilon_{\rm B,-3}^{\frac{p+2}{2(p+4)}} \varepsilon_{\rm e,-1}^{\frac{2(p-1)}{p+4}} A^{\frac{32+\alpha_s(p+6)}{2(p+4)(\alpha_s+8-2k)}}_{\rm st} \cr
&& \hspace{7.0cm} \times \,\tilde{E}_{50}^{\frac{4(p+2)-k(p+6)}{(p+4)(\alpha_s+8-2k)}}\,t^{-\frac{4(4+6p-\alpha_s)+k[4+6\alpha_s+p(\alpha_s-6)]}{2(p+4)(\alpha_s+8-2k)}}_4,\cr 
\nu_{\rm a,3} &\simeq& 2.13\times 10^{5}\,{\rm Hz}\,\left(\frac{1+z}{1.05}\right)^{-\frac{20+13\alpha_s-k(9\alpha_s+16)}{5(\alpha_s+8-2k)}} (1+Y)\,\varepsilon_{\rm B,-3}^{\frac{6}{5}} A^{\frac{44+9\alpha_s}{5(\alpha_s+8-2k)}}_{\rm st} \tilde{E}_{50}^{\frac{2(14-9k)}{5(\alpha_s+8-2k)}}t^{-\frac{20-8\alpha_s+3k(2+3\alpha_s)}{5(\alpha_s+8-2k)}}_4\,.
\eary
}

Table~\ref{Table2} shows the evolution of synchrotron light curves across the spectral and temporal indices, given the spectral breaks and maximum flux from eqs. \ref{En_br_syn_ism} and \ref{Self_abs_syn_ism}. Also presented in Table~\ref{Table3} are the closure relations for each cooling condition seen during the deceleration phase.

\paragraph{Post-jet-break decay phase}

During the post-jet-break decay phase, the Lorentz factor of the lowest energy electrons and the one of those with energy above which they cool effectively are
{\small
\bary\label{eLor_syn_ism_l}
\gamma_{\rm m}&=& 6.53\times 10^{1}\,\left(\frac{1+z}{1.05}\right)^{\frac{3-k}{\alpha_s+6-2k}} \zeta_{\rm e}^{-1}\, g(p)\,\beta^{-\frac{\alpha_s}{\alpha_s+6-2k}} \varepsilon_{\rm e, -1}\, A^{-\frac{1}{\alpha_s+6-2k}}_{\rm st}\, \tilde{E}_{50}^{\frac{1}{\alpha_s+6-2k}}\,t^{\frac{k-3}{\alpha_s+6-2k}}_{6},\cr
\gamma_{\rm c}&=& 3.47\times 10^{3}\,\left(\frac{1+z}{1.05}\right)^{\frac{\alpha_s+k-3-\alpha_s k}{\alpha_s+6-2k}} \beta^{\frac{\alpha_s(3-2k)}{\alpha_s+6-2k}} (1+Y)^{-1}\varepsilon_{\rm B,-3}^{-1} \, A^{-\frac{\alpha_s+3}{\alpha_s+6-2k}}_{\rm st}\, \tilde{E}^{\frac{2k-3}{\alpha_s+6-2k}}_{50}\,t^{\frac{3-\alpha_s+k(\alpha_s-1)}{\alpha_s+6-2k}}_6.
\eary
}
The characteristic and cooling synchrotron breaks and the maximum synchrotron flux become
{\small
\bary\label{En_br_syn_ism_l}
\nu_{\rm m}&\simeq& 1.33\times 10^{9}\,{\rm Hz}\,\left(\frac{1+z}{1.05}\right)^{\frac{12-2\alpha_s-4k+\alpha_s k}{2(\alpha_s+6-2k)}} \zeta_{\rm e}^{-2}g^2(p) \beta^{\frac{\alpha_s(k-4)}{\alpha_s+6-2k}}
\varepsilon_{\rm e,-1}^2 \varepsilon_{\rm B,-3}^{\frac{1}{2}} A^{\frac{\alpha_s-2}{2(\alpha_s+6-2k)}}_{\rm st} \tilde{E}_{50}^{\frac{4-k}{\alpha_s+6-2k}}t^{-\frac{24+(\alpha_s-8)k}{2(\alpha_s+6-2k)}}_6,\cr
\nu_{\rm c}&\simeq& 8.57\times 10^{14}\,{\rm Hz}\, \left(\frac{1+z}{1.05}\right)^{\frac{\alpha_s(2-3k)+4(k-3)}{2(\alpha_s+6-2k)}} (1+Y)^{-2}  \beta^{\frac{\alpha_s(4-3k)}{\alpha_s+6-2k}}\varepsilon_{\rm B,-3}^{-\frac{3}{2}}A^{-\frac{10+3\alpha_s}{2(\alpha_s+6-2k)}}_{\rm st}\tilde{E}_{50}^{\frac{3k-4}{\alpha_s+6-2k}}t^{\frac{\alpha_s(3k-4)}{2(\alpha_s+6-2k)}}_6,\cr 
F_{\rm max} &\simeq& 7.49\times 10^{-1}\,{\rm mJy}\,\left(\frac{1+z}{1.05}\right)^{\frac{36-2\alpha_s+3k(\alpha_s-4)}{2(\alpha_s+6-2k)}} \zeta_{\rm e}\,\beta^{\frac{\alpha_s(3k-8)}{\alpha_s+6-2k}} \varepsilon_{\rm B,-3}^{\frac{1}{2}}d^{-2}_{z,26.78}A^{\frac{3\alpha_s+2}{2(\alpha_s+6-2k)}}_{\rm st}\tilde{E}_{50}^{\frac{8-3k}{\alpha_s+6-2k}}t^{\frac{4(k-3)-3\alpha_s(k-2)}{2(\alpha_s+6-2k)}}_6\,.
\eary
}
The synchrotron spectral breaks in the self-absorption regime can be expressed as
{\small
\bary\label{Self_abs_syn_ism2}
\nu_{\rm a,1}&\simeq& 1.52\times 10^{2}\,{\rm Hz}\,\left(\frac{1+z}{1.05}\right)^{\frac{4k(\alpha_s+2)-8(\alpha_s+3)}{5(\alpha_s+6-2k)}}\zeta_{\rm e}\,g(p)^{-1}\beta^{\frac{8\alpha_s(k-1)}{5(\alpha_s+6-2k)}}\varepsilon_{\rm e,-1}^{-1} \varepsilon_{\rm B,-3}^{\frac{1}{5}} A^{\frac{4(\alpha_s+4)}{5(\alpha_s+6-2k)}}_{\rm st}\tilde{E}_{50}^{\frac{8(1-k)}{5(\alpha_s+6-2k)}}t^{\frac{3(\alpha_s-2)+2k(1-2\alpha_s)}{5(\alpha_s+6-2k)}}_6,\cr
\nu_{\rm a,2}&\simeq& 3.24\times 10^{5}\,{\rm Hz}\,\left(\frac{1+z}{1.05}\right)^{\frac{k[8+p(\alpha_s-4)+6\alpha_s]-2[p(\alpha_s-6)+6(\alpha_s+2)]}{2(p+4)(\alpha_s+6-2k)}}\zeta_{\rm e}^{-\frac{2(p-1)}{p+4}}g(p)^{\frac{2(p-1)}{p+4}}\beta^{\frac{\alpha_s[k(p+6)-4(p+2)]}{(p+4)(\alpha_s+6-2k)}}\varepsilon_{\rm B,-3}^{\frac{p+2}{2(p+4)}} \varepsilon_{\rm e,-1}^{\frac{2(p-1)}{p+4}} A^{\frac{20+p(\alpha_s-2)+6\alpha_s}{2(p+4)(\alpha_s+6-2k)}}_{\rm st}\cr
& &   \hspace{9cm}  \times \,\tilde{E}_{50}^{\frac{4(p+2)-k(p+6)}{(p+4)(\alpha_s+6-2k)}}\, t^{-\frac{4(6+6p-\alpha_s)+k[-8+p(\alpha_s-8)+6\alpha_s]}{2(p+4)(\alpha_s+6-2k)}}_6,\cr 
\nu_{\rm a,3} &\simeq& 1.61\times 10^{2}\,{\rm Hz}\,\left(\frac{1+z}{1.05}\right)^{\frac{6-13\alpha_s+k(9\alpha_s-2)}{5(\alpha_s+6-2k)}}(1+Y)\beta^{\frac{2\alpha_s(9k-14)}{5(\alpha_s+6-2k)}}\varepsilon_{\rm B,-3}^{\frac{6}{5}} A^{\frac{26+9\alpha_s}{5(\alpha_s+6-2k)}}_{\rm st} \tilde{E}_{50}^{\frac{2(14-9k)}{5(\alpha_s+6-2k)}}t^{-\frac{2[36-8\alpha_s+3k(3\alpha_s-4)]}{5(\alpha_s+6-2k)}}_6\,.
\eary
}
Table~\ref{Table2} shows the evolution of synchrotron light curves across the spectral and temporal indices, given the spectral breaks and maximum flux from eqs. \ref{En_br_syn_ism_l} and \ref{Self_abs_syn_ism2}. Also presented in Table~\ref{Table3} are the closure relations for each cooling condition seen during the post-jet-break decay phase.

\subsection{Sub-relativistic regime}

\paragraph{Coasting phase}
A constant velocity characterizes the coasting phase. For the lowest energy electrons and for those with energy above which they cool effectively, the Lorentz factors are

{\small
\bary\label{gamma_c}
\gamma_{\rm m}&=&  7.55\times10^{1}\,  g(p) \varepsilon_{\rm e,-1}\,\zeta_{\rm e}^{-1}\,\beta_{-0.22}^2,\,\cr
\gamma_{\rm c}&=& 2.15\times 10^{2}\,\,\,\left(\frac{1+z}{1.05}\right)^{1-k}(1+Y)^{-1}\,\varepsilon^{-1}_{\rm B,-3}\,A^{-1}_{\rm st} \beta_{-0.22}^{k-2}\,t_5^{k-1}\,.
\eary
}
The characteristic and cooling synchrotron breaks and the maximum synchrotron flux become
{\small
\bary\label{nu_syn}
\nu_{\rm m}&=&7.72\times 10^{9}\,{\rm Hz}\,\,g^2(p) \left(\frac{1+z}{1.05}\right)^{\frac{k-2}{2}}\zeta_{\rm e}^{-2}\varepsilon_{\rm e,-1}^2\,\varepsilon_{\rm B,-3}^\frac12 A^\frac{1}{2}_{\rm st} \beta_{-0.22}^\frac{10-k}{2}\,t_5^{-\frac{k}{2}},\cr
\nu_{\rm c}&=& 8.11\times 10^{11}\,{\rm Hz}\, \left(\frac{1+z}{1.05}\right)^{\frac{2-3k}{2}} (1+Y)^{-2}\,  \varepsilon_{\rm B,-3}^{-\frac32}\, A^{-\frac32}_{\rm st} \beta_{-0.22}^{\frac{3k-6}{2}}\,t_5^{\frac{3k-4}{2}},\cr
F_{\rm max}&=& 7.01\,{\rm mJy}\, \left(\frac{1+z}{1.05}\right)^{\frac{3k-4}{2}}\, \zeta_{\rm e}\,\varepsilon^{\frac12}_{\rm B,-3}\, d_{\rm z,26.78}^{-2}\, A^{\frac32}_{\rm st}\, \beta_{-0.22}^{\frac{8-3k}{2}}\,t_5^{\frac{3(2-k)}{2}}\,.
\eary
}

The synchrotron spectral breaks in the self-absorption regime can be expressed as
{\small
\bary\label{nu_syn_a}
\nu_{\rm a,1}&=& 3.21\times 10^{10}\,{\rm Hz}\,\left(\frac{1+z}{1.05}\right)^{\frac{4(k-2)}{5}}\zeta_{\rm e}\,g(p)^{-1}\, \varepsilon_{\rm e,-1}^{-1}\, \varepsilon_{\rm B,-3}^{\frac15}\, A^{\frac45}_{\rm st} \beta_{-0.22}^{-\frac{4k+5}{5}}\,t_5^{\frac{3-4k}{5}}\,,\cr
\nu^{\rm syn}_{\rm a,2}&=& 1.64\times 10^{10}\,{\rm Hz}\, \left(\frac{1+z}{1.05}\right)^{\frac{(k-2)(p+6)}{2(p+4)}}\zeta_{\rm e}^{-\frac{2(p-1)}{p+4}}g(p)^{\frac{2(p-1)}{p+4}}\, \varepsilon_{\rm e,-1}^{\frac{2(p-1)}{p+4}}\, \varepsilon_{\rm B,-3}^{\frac{p+2}{2(p+4)}} A^{\frac{p+6}{2(p+4)}}_{\rm st}\beta_{-0.22}^{ \frac{10p-kp-6k}{2(p+4)} }\,t_5^{\frac{4-kp-6k}{2(p+4)}},\cr
\nu_{\rm a,3}&=& 1.54\times 10^{12}\,{\rm Hz}\, \left(\frac{1+z}{1.05}\right)^{\frac{9k-13}{5}}\,(1+Y)\, \varepsilon_{\rm B,-3}^{\frac65}\, A^{\frac95}_{\rm st} \beta_{-0.22}^{\frac{15-9k}{5}}\,t_5^{\frac{8-9k}{5}}\,,
\eary
}

Table~\ref{Table2} shows the evolution of synchrotron light curves across the spectral and temporal indices, given the spectral breaks and maximum flux from eqs. \ref{En_br_syn_Coast} and \ref{Self_abs_syn_Coast}. Also presented in Table~\ref{Table3} are the closure relations for each cooling condition seen during the coasting phase.

\paragraph{Deceleration phase}
During the deceleration phase, the Lorentz factor of the lowest energy electrons and the one of those with energy above which they cool effectively are

{\small
\bary\label{gamma_dec}
\gamma_{\rm m}&=& 4.24\times 10^{1}\,{\rm Hz}\, \,\left(\frac{1+z}{1.05}\right)^{\frac{2(3-k)}{\alpha+5-k}}\,\zeta_{\rm e}^{-1}\,g(p)\, \varepsilon_{\rm e,-1}\, A^{-\frac{2}{\alpha_s+5-k}}_{\rm st}\,\tilde{E}_{50}^{\frac{2}{\alpha_s+5-k}}\, t_6^{\frac{2(k-3)}{\alpha+5-k}},\cr
\gamma_{\rm c}&=& 1.01\times 10^{2}\,{\rm Hz}\, \left(\frac{1+z}{1.05}\right)^{-\frac{k+1+\alpha(k-1)}{\alpha_s+5-k}}\, (1+Y)^{-1} \varepsilon^{-1}_{\rm B,-3}\, A^{-\frac{\alpha+3}{\alpha_s+5-k}}_{\rm st}\,\tilde{E}_{50}^{\frac{k-2}{\alpha_s+5-k}}\, t_6^{\frac{k+1+\alpha_s(k-1)}{\alpha_s+5-k}}\,.
\eary
}

The characteristic and cooling synchrotron breaks and the maximum synchrotron flux become
{\small
\bary\label{nu_syn_de}
\nu_{\rm m}&=& 1.68\times 10^{8}\,{\rm Hz}\,\,\left(\frac{1+z}{1.05}\right)^{\frac{20+k(\alpha_s-6)-2\alpha_s }{2(\alpha_s+5-k)}}\,\zeta_{\rm e}^{-2}\,g(p)^{2} \varepsilon^2_{\rm e,-1}\,\varepsilon^\frac12_{\rm B,-3}\,  A^{\frac{\alpha_s-5}{2(\alpha+5-k)}}_{\rm st}\, \tilde{E}_{50}^{\frac{10-k}{2(\alpha_s+5-k)}}\,t_6^{-\frac{30 + k(\alpha_s-8)}{2(\alpha+5-k)}},\cr
\nu_{\rm c}&=& 6.48\times 10^{11}\,{\rm Hz}\, \,\left(\frac{1+z}{1.05}\right)^{-\frac{8-2\alpha_s + k(3\alpha_s+2)}{2(\alpha_s +5-k)}}\, \varepsilon^{-\frac32}_{\rm B,-3}\, (1+Y)^{-2} \, A^{-\frac{3(\alpha_s+3)}{2(\alpha_s+5-k)}}_{\rm st}\tilde{E}_{50}^{\frac{3(k-2)}{2(\alpha_s+5-k)}}\,t_6^{\frac{k(4+3\alpha_s) - 4\alpha-2}{2(\alpha_s+5-k)}}\, ,\cr
F_{\rm \nu,max}&=& 5.2\,{\rm mJy}\,\,\left(\frac{1+z}{1.05}\right)^{\frac{4+2k-4\alpha+3k\alpha_s}{2(\alpha_s+5-k)}}\,\zeta_{\rm e}\, \varepsilon^{\frac12}_{\rm B,-3}\, d_{\rm z,26.78}^{-2}\, A^{\frac{3\alpha_s+7}{2(\alpha_s+5-k)}}_{\rm st}\,  \tilde{E}_{50}^{\frac{8-3k}{2(\alpha_s+5-k)}}\,t_6^{\frac{6-4k + 6\alpha_s-3k\alpha_s}{2(\alpha+5-k)}}.
\eary
}

In the self-absorption regime, the synchrotron break frequencies are 
{\small
\bary
\nu_{\rm a,1}&=& 7.49\times 10^{9}\,{\rm Hz}\, \left(\frac{1+z}{1.05}\right)^{-\frac{55+8\alpha_s-k(21+4\alpha)}{5(\alpha_s+5-k)}}\,\zeta_{\rm e}\,g(p)^{-1} \varepsilon_{\rm e,-1}^{-1} \varepsilon_{\rm B,-3}^{\frac15}\, A^{\frac{25+4\alpha}{5(\alpha_s+5-k)}}_{\rm st}\tilde{E}_{50}^{-\frac{4k+5}{5(\alpha+5-k)}}  t_6^{\frac{3\alpha_s-16k+30-4k\alpha}{ 5(\alpha_s+5-k)}}\, ,\cr
\nu_{\rm a,2}&=& 1.52\times 10^{9}\,{\rm Hz}\, \left(\frac{1+z}{1.05}\right)^{\frac{-2p(\alpha-10)+kp(\alpha_s-6)+6k(\alpha_s+4)-12(\alpha_s+5)}{2(p+4)(\alpha_s+5-k)}}\,\zeta_{\rm e}^{-\frac{2(p-1)}{p+4}}\, g(p)^{\frac{2(p-1)}{p+4}} \varepsilon_{\rm e,-1}^{\frac{2(p-1)}{p+4}}\, A^{\frac{\alpha_s p+6\alpha-5p+30}{2(p+4)(\alpha_s+5-k)}}_{\rm st} \varepsilon_{\rm B,-3}^{\frac{p+2}{2(p+4)}}\, \tilde{E}_{50}^{\frac{10p-kp-6k}{2(p+4)(\alpha_s+5-k)}}\cr
&&\hspace{9cm}\times\,\,t_6^{\frac{4\alpha_s-6k\alpha_s-\alpha_s kp+8kp-30p-16k+20}{2(p+4)(\alpha_s+5-k)}},\cr
\nu_{\rm a, 3}&=& 2.44\times 10^{9}\,{\rm Hz}\, \left(\frac{1+z}{1.05}\right)^{\frac{k(16+9\alpha_s)-13\alpha_s-20}{5(\alpha_s+5-k)}}\,(1+Y)\, \varepsilon_{\rm B,-3}^{\frac65}\, A^{\frac{3(3\alpha+10)}{ 5(\alpha_s+5-k)}}_{\rm st}\tilde{E}_{50}^{\frac{15-9k}{5(\alpha_s+5-k)}} t_6^{\frac{8\alpha_s-9k\alpha_s-11k-5}{5(\alpha_s+5-k)}}\,.
\eary
}
Table~\ref{Table2} shows the evolution of synchrotron light curves across the spectral and temporal indices, given the spectral breaks and maximum flux from eqs. \ref{En_br_syn_ism} and \ref{Self_abs_syn_ism}. Also presented in Table~\ref{Table3} are the closure relations for each cooling condition seen during the deceleration phase.


\bsp	
\label{lastpage}
\end{document}